
%
\def\unredoffs{}
\tolerance=1000\hfuzz=2pt
\catcode`\@=11 
\ifx\hyperdef\UNd@FiNeD\def\hyperdef#1#2#3#4{#4}\def\hyperref#1#2#3#4{#4}\def\href#1#2{#2}\fi
\magnification=1200\unredoffs\baselineskip=16pt plus 2pt minus 1pt
\def\Date#1{\vfill\leftline{#1}\tenpoint\supereject%
\footline={\hss\tenrm\hyperdef\hypernoname{page}\folio\folio\hss}}%

{\count255=\time\divide\count255 by 60 \xdef\hourmin{\number\count255}
 \multiply\count255 by-60\advance\count255 by\time
 \xdef\hourmin{\hourmin:\ifnum\count255<10 0\fi\the\count255}
}
\def\date{\number\day.\number\month.\number\year\ at \hourmin}


\def\nolabels{\def\wrlabeL##1{}\def\eqlabeL##1{}\def\reflabeL##1{}}
\def\writelabels{\def\wrlabeL##1{\leavevmode\vadjust{\rlap{\smash%
{\line{{\escapechar=` \hfill\rlap{\sevenrm\hskip.03in\string##1}}}}}}}%
\def\eqlabeL##1{{\escapechar-1\rlap{\sevenrm\hskip.05in\string##1}}}%
\def\reflabeL##1{\noexpand\llap{\noexpand\sevenrm\string\string\string##1}}}
\nolabels

\global\newcount\secno \global\secno=0
\global\newcount\meqno \global\meqno=1
\def\s@csym{}

\def\newsec#1\par{\global\advance\secno by1%
{\toks0{#1}\message{(\the\secno. \the\toks0)}}%
\global\subsecno=0\eqnres@t\let\s@csym\secsym\xdef\secn@m{\the\secno}\noindent
{\bf\hyperdef\hypernoname{section}{\the\secno}{\the\secno.} #1}%
\writetoca{{\string\hyperref{}{section}{\the\secno}{\bf \the\secno\quad}} {\bf #1}}\par%
\nobreak\medskip\nobreak\noindent\ignorespaces}
\def\eqnres@t{\xdef\secsym{\the\secno.}\global\meqno=1\bigbreak\bigskip}
\def\sequentialequations{\def\eqnres@t{\bigbreak}}\xdef\secsym{}

\global\newcount\subsecno \global\subsecno=0
\def\subsec#1\par{\global\advance\subsecno by1%
{\toks0{#1}\message{(\s@csym\the\subsecno. \the\toks0)}}%
\global\subsubsecno=0%
\ifnum\lastpenalty>9000\else\bigbreak\fi
\noindent{\it\hyperdef\hypernoname{subsection}{\secn@m.\the\subsecno}%
{\secn@m.\the\subsecno.} #1}\writetoca{\string\hskip1.45cm
{\string\hyperref{}{subsection}{\secn@m.\the\subsecno}{\secn@m.\the\subsecno.}}
{#1}}\par\nobreak\medskip\nobreak\noindent\ignorespaces}

\def\appendix#1#2{\global\meqno=1\global\subsecno=0\xdef\secsym{\hbox{#1.}}%
\bigbreak\bigskip\noindent{\bf Appendix \hyperdef\hypernoname{appendix}{#1}%
{#1.} #2}{\toks0{(#1. #2)}\message{\the\toks0}}%
\xdef\s@csym{#1.}\xdef\secn@m{#1}%
\writetoca{{\string\hyperref{}{appendix}{#1}{\bf {#1}\quad}} {\bf #2}}%
\par\nobreak\medskip\nobreak}

%
\def\checkm@de#1#2{\ifmmode{\def\f@rst##1{##1}\hyperdef\hypernoname{equation}%
{#1}{#2}}\else\hyperref{}{equation}{#1}{#2}\fi}
\def\eqnn#1{\DefWarn#1\xdef #1{(\noexpand\relax\noexpand\checkm@de%
{\s@csym\the\meqno}{\secsym\the\meqno})}%
\wrlabeL#1\writedef{#1\leftbracket#1}\global\advance\meqno by1}
\def\f@rst#1{\c@t#1a\em@ark}\def\c@t#1#2\em@ark{#1}
\def\eqna#1{\DefWarn#1\wrlabeL{#1$\{\}$}%
\xdef #1##1{(\noexpand\relax\noexpand\checkm@de%
{\s@csym\the\meqno\noexpand\f@rst{##1}1}{\hbox{$\secsym\the\meqno##1$}})}
\writedef{#1\numbersign1\leftbracket#1{\numbersign1}}\global\advance\meqno by1}
\def\eqn#1#2{\DefWarn#1%
\xdef #1{(\noexpand\hyperref{}{equation}{\s@csym\the\meqno}%
{\secsym\the\meqno})}$$#2\eqno(\hyperdef\hypernoname{equation}%
{\s@csym\the\meqno}{\secsym\the\meqno})\eqlabeL#1$$%
\writedef{#1\leftbracket#1}\global\advance\meqno by1}
\def\xeqn{\expandafter\xe@n}\def\xe@n(#1){#1}
\def\xeqna#1{\expandafter\xe@n#1}
\def\eqns#1{(\e@ns #1{\hbox{}})}
\def\e@ns#1{\ifx\UNd@FiNeD#1\message{eqnlabel \string#1 is undefined.}%
\xdef#1{(?.?)}\fi{\let\hyperref=\relax\xdef\next{#1}}%
\ifx\next\em@rk\def\next{}\else%
\ifx\next#1\xeqn#1\else\def\n@xt{#1}\ifx\n@xt\next#1\else\xeqna#1\fi
\fi\let\next=\e@ns\fi\next}

\def\DefWarn#1{\ifx\UNd@FiNeD#1\else
\immediate\write16{*** WARNING: the label \string#1 is already defined ***}\fi}
%
\newskip\footskip\footskip14pt plus 1pt minus 1pt 
\def\footnotefont{\ninepoint}\def\f@t#1{\footnotefont #1\@foot}
\def\f@@t{\baselineskip\footskip\bgroup\footnotefont\aftergroup\@foot\let\next}
\setbox\strutbox=\hbox{\vrule height9.5pt depth4.5pt width0pt}
\global\newcount\ftno \global\ftno=0
\def\foot{\global\advance\ftno by1\def\foot@rg{\hyperref{}{footnote}%
{\the\ftno}{\the\ftno}\xdef\foot@rg{\noexpand\hyperdef\noexpand\hypernoname%
{footnote}{\the\ftno}{\the\ftno}}}\footnote{$^{\foot@rg}$}}
%
%
%
\global\newcount\refno \global\refno=1
\newwrite\rfile
\def\ref{[\hyperref{}{reference}{\the\refno}{\the\refno}]\nref}
\def\nref#1{\DefWarn#1%
\xdef#1{[\noexpand\hyperref{}{reference}{\the\refno}{\the\refno}]}%
\writedef{#1\leftbracket#1}%
\ifnum\refno=1\immediate\openout\rfile=\jobname.refs\fi
\chardef\wfile=\rfile\immediate\write\rfile{\noexpand\item{[\noexpand\hyperdef%
\noexpand\hypernoname{reference}{\the\refno}{\the\refno}]\ }%
\reflabeL{#1\hskip.31in}\pctsign}\global\advance\refno by1\findarg}
\def\findarg#1#{\begingroup\obeylines\newlinechar=`\^^M\pass@rg}
{\obeylines\gdef\pass@rg#1{\writ@line\relax #1^^M\hbox{}^^M}%
\gdef\writ@line#1^^M{\expandafter\toks0\expandafter{\striprel@x #1}%
\edef\next{\the\toks0}\ifx\next\em@rk\let\next=\endgroup\else\ifx\next\empty%
\else\immediate\write\wfile{\the\toks0}\fi\let\next=\writ@line\fi\next\relax}}
\def\striprel@x#1{} \def\em@rk{\hbox{}}
\def\lref{\begingroup\obeylines\lr@f}
\def\lr@f#1#2{\DefWarn#1\gdef#1{\let#1=\UNd@FiNeD\ref#1{#2}}\endgroup\unskip}
\def\semi{;\hfil\break}
\def\addref#1{\immediate\write\rfile{\noexpand\item{}#1}} 
\def\listrefs{\vfill\supereject\immediate\closeout\rfile\writestoppt
\baselineskip=\footskip\centerline{{\bf References}}\bigskip{\parindent=20pt%
\frenchspacing\escapechar=` \input \jobname.refs\vfill\eject}\nonfrenchspacing}
\def\startrefs#1{\immediate\openout\rfile=\jobname.refs\refno=#1}
\def\xref{\expandafter\xr@f}\def\xr@f[#1]{#1}
\def\refs#1{\count255=1[\r@fs #1{\hbox{}}]}
\def\r@fs#1{\ifx\UNd@FiNeD#1\message{reflabel \string#1 is undefined.}%
\nref#1{need to supply reference \string#1.}\fi%
\vphantom{\hphantom{#1}}{\let\hyperref=\relax\xdef\next{#1}}%
\ifx\next\em@rk\def\next{}%
\else\ifx\next#1\ifodd\count255\relax\xref#1\count255=0\fi%
\else#1\count255=1\fi\let\next=\r@fs\fi\next}
%

%
\newwrite\ffile\global\newcount\figno \global\figno=1
\def\fig{fig.~\hyperref{}{figure}{\the\figno}{\the\figno}\nfig}
\def\nfig#1{\DefWarn#1%
\xdef#1{fig.~\noexpand\hyperref{}{figure}{\the\figno}{\the\figno}}%
\writedef{#1\leftbracket fig.\noexpand~\xfig#1}%
\ifnum\figno=1\immediate\openout\ffile=\jobname.figs\fi\chardef\wfile=\ffile%
{\let\hyperref=\relax
\immediate\write\ffile{\noexpand\medskip\noexpand\item{Fig.\ %
\noexpand\hyperdef\noexpand\hypernoname{figure}{\the\figno}{\the\figno}. }
\reflabeL{#1\hskip.55in}\pctsign}}\global\advance\figno by1\findarg}
\def\xfig{\expandafter\xf@g}\def\xf@g fig.\penalty\@M\ {}
\def\figs#1{figs.~\f@gs #1{\hbox{}}}
\def\f@gs#1{{\let\hyperref=\relax\xdef\next{#1}}\ifx\next\em@rk\def\next{}\else
\ifx\next#1\xfig #1\else#1\fi\let\next=\f@gs\fi\next}
%
\def\figin{\epsfcheck\figin}\def\figins{\epsfcheck\figins}
\def\epsfcheck{\ifx\epsfbox\UnDeFiNeD
\message{(NO epsf.tex, FIGURES WILL BE IGNORED)}
\gdef\figin##1{\vskip2in}\gdef\figins##1{\hskip.5in}
\else\message{(FIGURES WILL BE INCLUDED)}%
\gdef\figin##1{##1}\gdef\figins##1{##1}\fi}
\def\DefWarn#1{}
\def\figinsert{\goodbreak\topinsert}
\def\ifig#1#2#3{\DefWarn#1\xdef#1{fig.~\the\figno}
\writedef{#1\leftbracket fig.\noexpand~\the\figno}%
\figinsert\figin{\centerline{#3}}
\smallskip
\leftskip=0pt \rightskip=0pt
\baselineskip12pt\noindent
{{\bf Fig.~\the\figno}\ \ninepoint #2}
\medskip
\global\advance\figno by1\par\endinsert}
\newwrite\lfile
{\escapechar-1\xdef\pctsign{\string\%}\xdef\leftbracket{\string\{}
\xdef\rightbracket{\string\}}\xdef\numbersign{\string\#}}
\def\writedefs{\immediate\openout\lfile=label.defs \def\writedef##1{%
{\let\hyperref=\relax\let\hyperdef=\relax\let\hypernoname=\relax
 \immediate\write\lfile{\string\def\string##1\rightbracket}}}}%
\def\writestop{\def\writestoppt{\immediate\write\lfile{\string\pageno
 \the\pageno\string\startrefs\leftbracket\the\refno\rightbracket
 \string\def\string\secsym\leftbracket\secsym\rightbracket
 \string\secno\the\secno\string\meqno\the\meqno}\immediate\closeout\lfile}}
\def\writestoppt{}\def\writedef#1{}

\def\seclab#1{\DefWarn#1%
\xdef #1{\noexpand\hyperref{}{section}{\the\secno}{\the\secno}}%
\writedef{#1\leftbracket#1}\wrlabeL{#1=#1}}
\def\subseclab#1{\DefWarn#1%
\xdef #1{\noexpand\hyperref{}{subsection}{\the\secno.\the\subsecno}%
{\the\secno.\the\subsecno}}\writedef{#1\leftbracket#1}\wrlabeL{#1=#1}}
\def\applab#1{\DefWarn#1%
\xdef #1{\noexpand\hyperref{}{appendix}{\secn@m}{\secn@m}}%
\writedef{#1\leftbracket#1}\wrlabeL{#1=#1}}
\newwrite\tfile \def\writetoca#1{}
\def\leaderfill{\leaders\hbox to 1em{\hss.\hss}\hfill}
\def\writetoc{\immediate\openout\tfile=\jobname.toc
   \def\writetoca##1{{\edef\next{\write\tfile{\noindent ##1
   \string\leaderfill{
   \string\hyperref{}{page}{\noexpand\number\pageno}%
   {\noexpand\number\pageno}} \par}}\next}}
}
\newread\ch@ckfile
\def\listtoc{\immediate\closeout\tfile\immediate\openin\ch@ckfile=\jobname.toc
\ifeof\ch@ckfile\message{no file \jobname.toc, no table of contents this pass}%
\else\closein\ch@ckfile\centerline{\bf Contents}\nobreak\medskip%
{\baselineskip=16pt\footnotefont\parskip=0pt\catcode`\@=11\input\jobname.toc
\catcode`\@=12\bigbreak\bigskip}\fi}
\catcode`\@=12 
\def\tenpoint{\def\rm{\fam0\tenrm}
\textfont0=\tenrm \scriptfont0=\sevenrm \scriptscriptfont0=\fiverm
\textfont1=\teni  \scriptfont1=\seveni  \scriptscriptfont1=\fivei
\textfont2=\tensy \scriptfont2=\sevensy \scriptscriptfont2=\fivesy
\textfont\itfam=\tenit \def\it{\fam\itfam\tenit}\def\footnotefont{\ninepoint}%
\textfont\bffam=\tenbf \def\bf{\fam\bffam\tenbf}\def\sl{\fam\slfam\tensl}\rm}
\font\ninerm=cmr9 \font\sixrm=cmr6 \font\ninei=cmmi9 \font\sixi=cmmi6
\font\ninesy=cmsy9 \font\sixsy=cmsy6 \font\ninebf=cmbx9
\font\nineit=cmti9 \font\ninesl=cmsl9 \skewchar\ninei='177
\skewchar\sixi='177 \skewchar\ninesy='60 \skewchar\sixsy='60
\def\ninepoint{\def\rm{\fam0\ninerm}
\textfont0=\ninerm \scriptfont0=\sixrm \scriptscriptfont0=\fiverm
\textfont1=\ninei \scriptfont1=\sixi \scriptscriptfont1=\fivei
\textfont2=\ninesy \scriptfont2=\sixsy \scriptscriptfont2=\fivesy
\textfont\itfam=\ninei \def\it{\fam\itfam\nineit}\def\sl{\fam\slfam\ninesl}%
\textfont\bffam=\ninebf \def\bf{\fam\bffam\ninebf}\rm}
%
\hyphenation{anom-aly anom-alies coun-ter-term coun-ter-terms}

\global\newcount\subsubsecno \global\subsubsecno=0
\def\subsubsec#1\par{\global\advance\subsubsecno by1%
{\toks0{#1}\message{(\the\secno\the\subsecno\the\subsubsecno. \the\toks0)}}%
\ifnum\lastpenalty>9000\else\bigbreak\fi
\noindent{\it\hyperdef\hypernoname{subsubsection}{\the\secno.\the\subsecno\the\subsubsecno}%
{\the\secno.\the\subsecno.\the\subsubsecno.} #1}
\par\nobreak\medskip\nobreak\noindent\ignorespaces}

\def\DefWarn#1{}
\def\tikzcaption#1#2{\DefWarn#1\xdef#1{Fig.~\the\figno}
\writedef{#1\leftbracket Fig.\noexpand~\the\figno}%
{
\smallskip
\leftskip=20pt \rightskip=20pt \baselineskip12pt\noindent
{{\bf Fig.~\the\figno}\ \ninepoint #2}
\bigskip
\global\advance\figno by1 \par}}

\def\ntoalpha#1{%
\ifcase#1%
@%
\or A\or B\or C\or D\or E\or F\or G\or H\or I
\fi
}

\global\newcount\appno \global\appno=1
\def\applab#1{\xdef #1{\ntoalpha\appno}\writedef{#1\leftbracket#1}\wrlabeL{#1=#1}
\global\advance\appno by1}

\def\preprint#1 #2\par{\rightline{\vbox{\baselineskip12pt\hbox{#1}\hbox{#2}}}\vskip2cm}
%
\def\title#1\par{\centerline{\bf #1}\nopagenumbers\pageno=0}
\def\author#1\par{\bigskip\bigskip\centerline{#1}}

\newcount\addressno

\def\email#1#2{\unskip$^#1$\footnote{\null}{\kern-\parindent \llap{$^#1$\hskip1pt}email: #2}}

\def\startcenter{%
  \par
  \begingroup
  \leftskip=0pt plus 1fil
  \rightskip=\leftskip
  \parindent=0pt
  \parfillskip=0pt
}
\def\stopcenter{\endgroup}

\def\address{\bigskip%
  \ifnum\the\addressno=0\else\stopcenter\endgroup\fi
  \advance\addressno by 1%
  \begingroup
  \startcenter
  \it
  \obeylines
  \addressAux
}
\def\addressAux#1{#1}

\def\abstract{\stopcenter\endgroup\bigskip\bigskip\noindent}

\def\Dsl{\,\raise.15ex\hbox{/}\mkern-13.5mu D} 
\def\dsl{\raise.15ex\hbox{/}\kern-.57em\partial}
 
\def\boxeqn#1{\vcenter{\vbox{\hrule\hbox{\vrule\kern3pt\vbox{\kern3pt
	\hbox{${\displaystyle #1}$}\kern3pt}\kern3pt\vrule}\hrule}}}


\def\ap{{\alpha^{\prime}}}

\def\a{\alpha}
\def\b{{\beta}}
\def\g{{\gamma}}
\def\d{{\delta}}

\def\l{\lambda}

\def\t{{\theta}}

\def\half{{1\over 2}}

\def\bar{\overline}
\def\({\left(}
\def\){\right)}



\def\qed{\hbox{\hskip 3pt
\vbox{\hrule\hbox to 7pt{\vrule height 7pt\hfill\vrule}
\hrule}}\hskip3pt}

\overfullrule=0pt\relax

\frenchspacing

\newread\instream \openin\instream= label.defs
\ifeof\instream \message{No labels in advance yet. Wait till next pass.}
\else \closein\instream \input label.defs
\fi
\writedefs

\def\arXiv:#1].{\hepthStrip#1 \nil}
\def\hepthStrip#1 #2\nil{\href{http://arxiv.org/abs/#1}{arXiv:#1 #2\unskip}].}

\input epsf.sty
\input amssym.tex 


\preprint DAMTP--2015--25

\title Two-loop five-point amplitudes of super Yang--Mills and supergravity\par
\title in pure spinor superspace

\author Carlos R. Mafra\email{\dagger}{c.r.mafra@damtp.cam.ac.uk} and Oliver Schlotterer\email{\star}{olivers@aei.mpg.de}

\address
$^\dagger$DAMTP, University of Cambridge
Wilberforce Road, Cambridge, CB3 0WA, UK

\address
$^\star$Max--Planck--Institut f\"ur Gravitationsphysik
Albert--Einstein--Institut, 14476 Potsdam, Germany

\abstract
Supersymmetric integrands for the two-loop five-point amplitudes in ten-dimensional super
Yang--Mills and type II supergravity are proposed. The kinematic numerators are manifestly local and
satisfy the duality between color and kinematics described by Bern, Carrasco and Johansson. Our results
are expected to reproduce the integrated two-loop amplitudes in dimensions $D<7$. The UV divergence in
the critical dimension $D=7$ matches the low-energy limit of the corresponding superstring amplitudes
and is written in terms of SYM tree amplitudes.

\Date {May 2015}

\newif\iffig
\figfalse


\lref\HullTownsend{
	C.~M.~Hull and P.~K.~Townsend,
	``Unity of superstring dualities,''
	Nucl.\ Phys.\ B {\bf 438}, 109 (1995).
	[hep-th/9410167].
}

\lref\Rviolating{
	M.B.~Green, M.~Gutperle and H.~h.~Kwon,
	``Sixteen fermion and related terms in M theory on $T^2$,''
	Phys.\ Lett.\ B {\bf 421}, 149 (1998).
	[hep-th/9710151].
\semi
	M.B.~Green,
	``Interconnections between type II superstrings, M theory and N=4 supersymmetric Yang-Mills,''
	Lect.\ Notes Phys.\  {\bf 525}, 22 (1999).
	[hep-th/9903124].
}

\lref\GrossSloan{
	D.J.~Gross and J.~H.~Sloan,
	``The Quartic Effective Action for the Heterotic String,''
	Nucl.\ Phys.\ B {\bf 291}, 41 (1987).
}

\lref\WWW{
	C.R.~Mafra and O.~Schlotterer,
	{\tt http://www.damtp.cam.ac.uk/user/crm66/SYM/pss.html}
}

\lref\CarrascoYPA{
  J.~J.~M.~Carrasco, R.~Kallosh, R.~Roiban and A.~A.~Tseytlin,
  ``On the U(1) duality anomaly and the S-matrix of N=4 supergravity,''
JHEP {\bf 1307}, 029 (2013).
[arXiv:1303.6219 [hep-th]].
}

\lref\tropical{
	P.~Tourkine,
	``Tropical Amplitudes,''
	[arXiv:1309.3551 [hep-th]].
}

\lref\BernTwoLoop{
	Z.~Bern, L.~J.~Dixon, D.~C.~Dunbar, M.~Perelstein and J.~S.~Rozowsky,
	``On the relationship between Yang-Mills theory and gravity and its implication for ultraviolet divergences,''
	Nucl.\ Phys.\ B {\bf 530}, 401 (1998).
	[hep-th/9802162].
}

\lref\BernUF{
	Z.~Bern, J.~J.~M.~Carrasco, L.~J.~Dixon, H.~Johansson and R.~Roiban,
	``Simplifying Multiloop Integrands and Ultraviolet Divergences of Gauge Theory and Gravity Amplitudes,''
	Phys.\ Rev.\ D {\bf 85}, 105014 (2012).
	[arXiv:1201.5366 [hep-th]].
}

\lref\dixon{
	L.J.~Dixon,
	``Calculating scattering amplitudes efficiently,''
	In *Boulder 1995, QCD and beyond* 539-582.
	[hep-ph/9601359].
}

\lref\bigHowe{
	P.S.~Howe,
	``Pure Spinors Lines In Superspace And Ten-Dimensional Supersymmetric
	Theories,''
	Phys.\ Lett.\  B {\bf 258}, 141 (1991)
	[Addendum-ibid.\  B {\bf 259}, 511 (1991)].
	\semi
	P.S.~Howe,
	``Pure Spinors, Function Superspaces And Supergravity Theories In
	Ten-Dimensions And Eleven-Dimensions,''
	Phys.\ Lett.\  B {\bf 273}, 90 (1991).
}

\lref\GreenFT{
	M.B.~Green, J.H.~Schwarz and L.~Brink,
	``N=4 Yang-Mills and N=8 Supergravity as Limits of String Theories,''
	Nucl.\ Phys.\ B {\bf 198}, 474 (1982).
}

\lref\MafraGJA{
	C.R.~Mafra and O.~Schlotterer,
	``Towards one-loop SYM amplitudes from the pure spinor BRST cohomology,''
	Fortsch.\ Phys.\  {\bf 63}, no. 2, 105 (2015).
	[arXiv:1410.0668 [hep-th]].
}
\lref\BernZX{
	Z.~Bern, L.~J.~Dixon, D.~C.~Dunbar and D.~A.~Kosower,
	``One loop n point gauge theory amplitudes, unitarity and collinear limits,''
	Nucl.\ Phys.\ B {\bf 425}, 217 (1994).
	[hep-ph/9403226].
}

\lref\BCJloop{
	Z.~Bern, J.~J.~M.~Carrasco and H.~Johansson,
	``Perturbative Quantum Gravity as a Double Copy of Gauge Theory,''
	Phys.\ Rev.\ Lett.\  {\bf 105}, 061602 (2010).
	[arXiv:1004.0476 [hep-th]].
}
\lref\yutin{
	Z.~Bern, T.~Dennen, Y.~t.~Huang and M.~Kiermaier,
	``Gravity as the Square of Gauge Theory,''
	Phys.\ Rev.\ D {\bf 82}, 065003 (2010).
	[arXiv:1004.0693 [hep-th]].
}

\lref\BjerrumBohrHN{
	N.~E.~J.~Bjerrum-Bohr, P.~H.~Damgaard, T.~Sondergaard and P.~Vanhove,
	``The Momentum Kernel of Gauge and Gravity Theories,''
	JHEP {\bf 1101}, 001 (2011).
	[arXiv:1010.3933 [hep-th]].
}

\lref\BCJ{
	Z.~Bern, J.J.M.~Carrasco and H.~Johansson,
	``New Relations for Gauge-Theory Amplitudes,''
	Phys.\ Rev.\ D {\bf 78}, 085011 (2008).
	[arXiv:0805.3993 [hep-ph]].
}

\lref\MafraKJ{
	C.R.~Mafra, O.~Schlotterer and S.~Stieberger,
	``Explicit BCJ Numerators from Pure Spinors,''
	JHEP {\bf 1107}, 092 (2011).
	[arXiv:1104.5224 [hep-th]].
}

\lref\oneloopbb{
	C.R.~Mafra and O.~Schlotterer,
	``The Structure of n-Point One-Loop Open Superstring Amplitudes,''
	JHEP {\bf 1408}, 099 (2014).
	[arXiv:1203.6215 [hep-th]].
}

\lref\MafraGSA{
	C.R.~Mafra and O.~Schlotterer,
	``Cohomology foundations of one-loop amplitudes in pure spinor superspace,''
	[arXiv:1408.3605 [hep-th]].
}

\lref\superpoincare{
	N.~Berkovits,
	``Super-Poincare covariant quantization of the superstring,''
	JHEP {\bf 0004}, 018 (2000)
	[arXiv:hep-th/0001035].
}
\lref\multiloop{
	N.~Berkovits,
	``Multiloop amplitudes and vanishing theorems using the pure spinor
	formalism for the superstring,''
	JHEP {\bf 0409}, 047 (2004)
	[arXiv:hep-th/0406055].
}

\lref\StiebergerHQ{
	S.~Stieberger,
	``Open \& Closed vs. Pure Open String Disk Amplitudes,''
	[arXiv:0907.2211 [hep-th]].
}

\lref\BjerrumBohrRD{
	N.~E.~J.~Bjerrum-Bohr, P.~H.~Damgaard and P.~Vanhove,
	``Minimal Basis for Gauge Theory Amplitudes,''
	Phys.\ Rev.\ Lett.\  {\bf 103}, 161602 (2009).
	[arXiv:0907.1425 [hep-th]].
}

\lref\NMPS{
	N.~Berkovits,
	``Pure spinor formalism as an N = 2 topological string,''
	JHEP {\bf 0510}, 089 (2005)
	[arXiv:hep-th/0509120].
}

\lref\PSanomaly{
	N.~Berkovits and C.~R.~Mafra,
	``Some superstring amplitude computations with the non-minimal pure spinor
	formalism,''
	JHEP {\bf 0611}, 079 (2006)
	[arXiv:hep-th/0607187].
}
\lref\FORM{
	J.A.M.~Vermaseren,
	``New features of FORM,''
	[arXiv:math-ph/0010025].
}

\lref\twoloop{
	N.~Berkovits,
	``Super-Poincare covariant two-loop superstring amplitudes,''
	JHEP {\bf 0601}, 005 (2006).
	[hep-th/0503197].
}

\lref\coefftwo{
	H.~Gomez, C.R.~Mafra,
	``The Overall Coefficient of the Two-loop Superstring Amplitude Using Pure Spinors,''
	JHEP {\bf 1005}, 017 (2010).
	[arXiv:1003.0678 [hep-th]].
}
\lref\thetaSYM{
  	J.P.~Harnad and S.~Shnider,
	``Constraints And Field Equations For Ten-Dimensional Superyang-Mills
  	Theory,''
  	Commun.\ Math.\ Phys.\  {\bf 106}, 183 (1986).
\semi
	P.A.~Grassi and L.~Tamassia,
	``Vertex operators for closed superstrings,''
	JHEP {\bf 0407}, 071 (2004)
	[arXiv:hep-th/0405072].
\semi
	G.~Policastro and D.~Tsimpis,
	``$R^4$, purified,''
	Class.\ Quant.\ Grav.\  {\bf 23}, 4753 (2006).
	[arXiv:hep-th/0603165].
}

\lref\BrinkBC{
	L.~Brink, J.H.~Schwarz and J.~Scherk,
	``Supersymmetric Yang-Mills Theories,''
	Nucl.\ Phys.\ B {\bf 121}, 77 (1977)..
}

\lref\refSYM{
	E.Witten,
	``Twistor-Like Transform In Ten-Dimensions''
	Nucl.Phys. B {\bf 266}, 245~(1986)
}

\lref\treebbI{
	C.R.~Mafra, O.~Schlotterer and S.~Stieberger,
	``Complete N-Point Superstring Disk Amplitude I. Pure Spinor Computation,'' Nucl.\ Phys.\ B {\bf 873}, 419 (2013).
	[arXiv:1106.2645 [hep-th]].
\semi
	C.R.~Mafra, O.~Schlotterer and S.~Stieberger,
	``Complete N-Point Superstring Disk Amplitude II. Amplitude and Hypergeometric Function Structure,'' Nucl.\ Phys.\ B {\bf 873}, 461 (2013).
	[arXiv:1106.2646 [hep-th]].
}

\lref\CarrascoMN{
	J.J.~Carrasco and H.~Johansson,
	``Five-Point Amplitudes in N=4 Super-Yang-Mills Theory and N=8 Supergravity,''
	Phys.\ Rev.\ D {\bf 85}, 025006 (2012).
	[arXiv:1106.4711 [hep-th]].
}

\lref\PSS{
	C.R.~Mafra,
	``PSS: A FORM Program to Evaluate Pure Spinor Superspace Expressions,''
	[arXiv:1007.4999 [hep-th]].
}
\lref\mafraids{
	C.R.~Mafra,
	``Pure Spinor Superspace Identities for Massless Four-point Kinematic Factors,''
	JHEP {\bf 0804}, 093 (2008).
	[arXiv:0801.0580 [hep-th]].
}
\lref\towards{
	C.R.~Mafra,
	``Towards Field Theory Amplitudes From the Cohomology of Pure Spinor Superspace,''
	JHEP {\bf 1011}, 096 (2010).
	[arXiv:1007.3639 [hep-th]].
}
\lref\threeloop{
	H.~Gomez and C.R.~Mafra,
	``The closed-string 3-loop amplitude and S-duality,''
	JHEP {\bf 1310}, 217 (2013).
	[arXiv:1308.6567 [hep-th]].
}

\lref\motivic{
	O.~Schlotterer and S.~Stieberger,
	``Motivic Multiple Zeta Values and Superstring Amplitudes,''
	J.\ Phys.\ A {\bf 46}, 475401 (2013).
	[arXiv:1205.1516 [hep-th]].
}

\lref\OchirovXBA{
	A.~Ochirov and P.~Tourkine,
	``BCJ duality and double copy in the closed string sector,''
	JHEP {\bf 1405}, 136 (2014).
	[arXiv:1312.1326 [hep-th]].
}

\lref\GreenYU{
	M.B.~Green, J.G.~Russo and P.~Vanhove,
	``Ultraviolet properties of maximal supergravity,''
	Phys.\ Rev.\ Lett.\  {\bf 98}, 131602 (2007).
	[hep-th/0611273].
}

\lref\FiveSdual{
	M.B.~Green, C.R.~Mafra and O.~Schlotterer,
	``Multiparticle one-loop amplitudes and S-duality in closed superstring theory,''
	JHEP {\bf 1310}, 188 (2013).
	[arXiv:1307.3534 [hep-th]].
}
\lref\EOMBBs{
	C.R.~Mafra and O.~Schlotterer,
	``Multiparticle SYM equations of motion and pure spinor BRST blocks,''
	JHEP {\bf 1407}, 153 (2014).
	[arXiv:1404.4986 [hep-th]].
}
\lref\HighSYM{
	C.R.~Mafra and O.~Schlotterer,
	``A solution to the non-linear equations of D=10 super Yang-Mills theory,''
[arXiv:1501.05562 [hep-th]].
}

\lref\BilalHB{
	A.~Bilal,
	``Higher derivative corrections to the nonAbelian Born-Infeld action,''
	Nucl.\ Phys.\ B {\bf 618}, 21 (2001).
	[hep-th/0106062].
}

\lref\MedinaNK{
	R.~Medina, F.~T.~Brandt and F.~R.~Machado,
	``The Open superstring five point amplitude revisited,''
	JHEP {\bf 0207}, 071 (2002).
	[hep-th/0208121].
}

\lref\twolooptwo{
	N.~Berkovits and C.R.~Mafra,
	``Equivalence of two-loop superstring amplitudes in the pure spinor and RNS formalisms,''
	Phys.\ Rev.\ Lett.\  {\bf 96}, 011602 (2006).
	[hep-th/0509234].
}

\lref\WWWalpha{
	J.~Br\"odel, O.~Schlotterer, S.~Stieberger,
{\tt http://mzv.mpp.mpg.de}
}

\lref\MafraJQ{
	C.R.~Mafra, O.~Schlotterer, S.~Stieberger and D.~Tsimpis,
	``A recursive method for SYM n-point tree amplitudes,''
	Phys.\ Rev.\ D {\bf 83}, 126012 (2011).
	[arXiv:1012.3981 [hep-th]].
}

\lref\verlinde{
	E.P.~Verlinde and H.L.~Verlinde,
	``Chiral bosonization, determinants and the string partition function,''
	Nucl.\ Phys.\  B {\bf 288}, 357 (1987).
}

\lref\siegel{
	W.~Siegel,
	``Classical Superstring Mechanics,''
	Nucl.\ Phys.\  {\bf B263}, 93 (1986).
}

\lref\KLTref{
	H.~Kawai, D.C.~Lewellen and S.H.H.~Tye,
	``A Relation Between Tree Amplitudes of Closed and Open Strings,''
	Nucl.\ Phys.\ B {\bf 269}, 1 (1986).
}

\lref\AdamoHOA{
	T.~Adamo and E.~Casali,
	``Scattering equations, supergravity integrands, and pure spinors,''
	[arXiv:1502.06826 [hep-th]].
}

\lref\twoloopfive{
	H.~Gomez, C.R.~Mafra and O.~Schlotterer,
	``The two-loop superstring five-point amplitude and S-duality,''
	[arXiv:1504.02759 [hep-th]].
}

\lref\BernTQ{
	Z.~Bern, J.~J.~M.~Carrasco, L.~J.~Dixon, H.~Johansson and R.~Roiban,
	``The Complete Four-Loop Four-Point Amplitude in N=4 Super-Yang-Mills Theory,''
	Phys.\ Rev.\ D {\bf 82}, 125040 (2010).
	[arXiv:1008.3327 [hep-th]].
}

\lref\MarcusEI{
	N.~Marcus and A.~Sagnotti,
	``The Ultraviolet Behavior of $N=4$ {Yang-Mills} and the Power Counting of Extended Superspace,''
	Nucl.\ Phys.\ B {\bf 256}, 77 (1985)..
}

\lref\stieFive{
	S.~Stieberger and T.~R.~Taylor,
	``Multi-Gluon Scattering in Open Superstring Theory,''
	Phys.\ Rev.\ D {\bf 74}, 126007 (2006).
	[hep-th/0609175].
}

\lref\medina{
	L.~A.~Barreiro and R.~Medina,
	``5-field terms in the open superstring effective action,''
	JHEP {\bf 0503}, 055 (2005).
	[hep-th/0503182].
}

\listtoc
\writetoc
\filbreak

\newsec{Introduction}

Tree-level and one-loop scattering amplitudes of ten-dimensional super Yang--Mills (SYM) have been
recently determined using a method based on two fundamental principles: locality and BRST invariance
\refs{\towards, \MafraJQ, \MafraGJA}. Locality refers to the expansion of amplitudes in terms of cubic
graphs with definite propagator structure \refs{\BCJ,\BCJloop}, and BRST invariance is a property of pure spinor
superspace that guarantees manifest supersymmetry and gauge invariance \superpoincare. Even though the
notion of BRST invariance is motivated by the pure spinor formalism of the superstring \superpoincare,
pure spinor variables are long known to simplify the description of ten-dimensional SYM \bigHowe, as
will be corroborated once more by this paper.

In the subsequent, we will describe the two-loop extension of this method and use it to derive the
five-point two-loop integrand of ten-dimensional SYM in an intuitive manner. In doing so, we follow
closely the organization found in a beautiful paper by Carrasco and Johansson \CarrascoMN\ which makes
the symmetry between color and kinematic degrees of freedom \refs{\BCJ,\BCJloop} manifest and thereby
leads to the supergravity integrand without any extra effort \yutin. As an additional benefit of the
pure spinor superspace representation, the kinematic numerators are manifestly local due to the very nature
of our method (bypassing the inverse Gram determinants in \CarrascoMN) and do not require any
constructive input from unitarity.

As the main result of this paper, the color-dressed five-point two-loop amplitude of SYM in ten
dimensions will be explicitly constructed in \delttt, following the guidelines of the method described
in section~3. The polarization dependence is furnished by two local kinematic building blocks
$T_{12,3|4,5}$ and $T^m_{1,2,3|4,5}$ written in pure spinor superspace and inspired by string theory
whose bosonic components can be downloaded from \WWW. Thanks to their compatibility with the
Bern--Carrasco--Johansson (BCJ) duality between color and kinematics \refs{\BCJ,\BCJloop}, the
corresponding supergravity amplitude is a straightforward corollary \yutin\ and given in \sugratwo.

Since our results are formulated in ten dimensions, standard dimensional reduction gives rise to their
lower-dimensional counterparts \BrinkBC. In section~\secfour, the UV divergences of maximally supersymmetric SYM and
supergravity in the two-loop critical dimension $D=7$ are written in terms of SYM tree amplitudes.
As a consistency check \GreenYU, the superspace expression for the supergravity UV divergence in \UVfive\ matches the
low-energy limit of the closed-string amplitude \twoloopfive.
However, type IIB superstring two-loop amplitudes violate the $U(1)$ R-symmetry as a consequence of
S-duality \refs{\HullTownsend,\Rviolating,\twoloopfive} while supergravity and its two-loop UV
divergence in $D=7$ dimensions conserve it.
This apparent paradox is
resolved by the prefactor $(7-D)$ in the R-symmetry violating components that appears once
the ten-dimensional superstring kinematic factor is dimensionally reduced to involve a $D$-dimensional
dilaton state. The same mechanism applies to the one-loop UV divergence,
where a prefactor of $(8-D)$ along with a $D$-dimensional dilaton reconciles
S-duality properties of the superstring amplitude \refs{\FiveSdual,\twoloopfive} with
R-symmetry of supergravity, see appendix~\appUV.

\newsec{Review}
\par\seclab\secone

\subsec BCJ duality between color and kinematics
\par\subseclab\secBCJ

\ifig\figBCJ{The Jacobi identity implies the vanishing of the color factors associated to a triplet of cubic
graphs, $C_i + C_j + C_k = 0$. In the above diagrams, the legs
$a$, $b$, $c$ and $d$ may represent arbitrary subdiagrams. The BCJ duality
states that their corresponding kinematic numerators $N_i(\ell)$ can be chosen such that $N_i(\ell) +
N_j(\ell) + N_k(\ell) = 0$.}
{\epsfxsize=0.70\hsize\epsfbox{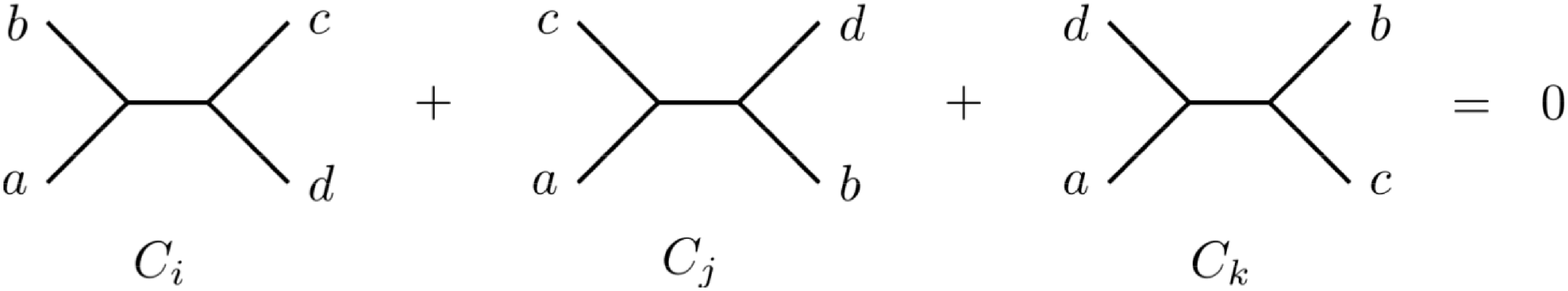}}

\noindent Bern, Carrasco and Johansson (BCJ) proposed an organization scheme for gauge theory
amplitudes based on cubic vertices where color and
kinematic degrees of freedom enter on completely symmetric footing \refs{\BCJ,\BCJloop}:
\eqn\cubicB{
{\cal A}^{g-{\rm loop}}_n = \int \prod_{j=1}^g {d^D \ell_j \over (2\pi)^D} \sum_{\Gamma_i} {N_i (\ell) C_i \over \prod_k P_{k,i}(\ell)}
}
The sum is understood to encompass all ``cubic'' graphs $\Gamma_i$ with $n$ external legs, $g$ loops
and only trivalent vertices as well as appropriate symmetry factors to avoid overcounting. The
propagators $P_{k,i}(\ell)$ refer to the squared momentum in the $k^{\rm th}$ internal line of the
$i^{\rm th}$ graph. The color tensors $C_i$ are obtained by dressing each vertex of $\Gamma_i$ with a
factor of $f^{abc}$, the structure constants of the gauge group, and each internal line by
$\delta^{ab}$. Finally, the kinematic numerators $N_i(\ell)$ encode the dependence on polarizations and
(external or internal) momenta $k$, $\ell$. They furnish the only ingredient of \cubicB\ that cannot
be immediately read off from the graphs, and they will be in the main focus of this work.

Triplets of color tensors $C_i,C_j,C_k$ associated 
with diagrams $i,j,k$ that differ in only one propagator (see \figBCJ) vanish due to the Jacobi identity 
\eqn\coljac{
f^{abe} f^{cde} +  f^{bce}  f^{ade} + f^{cae}  f^{bde}  =  0 
}
valid for any gauge group. The BCJ conjecture \refs{\BCJ, \BCJloop} states that amplitudes in \cubicB\
can be represented such that for any vanishing color triplet $C_i + C_j + C_k$, the corresponding
kinematic decorations $N_i(\ell)+ N_j(\ell) + N_k(\ell)$ of diagrams $i,j,k$ vanish as well. This
statement is illustrated in \figBCJ\ and understood to hold for any value of the loop momenta $\ell$.
Of course, the momentum in the four external lines represented by $a, b, c, d$ and the subdiagrams beyond
them have to be the same in the three graphs $i,j,k$.


\noindent
Once a gauge theory amplitude \cubicB\ has been cast into such a ``BCJ form'',
\eqn\BCJform{
C_i + C_j + C_k = 0
\ \ \Rightarrow \ \
N_i(\ell)+ N_j(\ell) + N_k(\ell)=0 \ ,
}
then the corresponding gravity amplitude follows from trading color tensors for a second copy of the kinematic
numerators, $C_i \rightarrow \tilde N_i(\ell)$ \yutin
\eqn\cubicC{
{\cal M}^{g-{\rm loop}}_n = \int \prod_{j=1}^g {d^D \ell_j \over (2\pi)^D} \sum_{\Gamma_i} {N_i (\ell) \tilde N_i(\ell) \over \prod_k P_{k,i}(\ell)} \ .
}
The second copy of $\tilde N_i(\ell)$ does not need to stem from the same gauge theory or satisfy the
kinematic Jacobi identities \BCJform. The polarization tensors of the resulting gravity 
amplitudes \cubicC\ are then given by the tensor products of the gauge theory polarizations 
contained in $N_i(\ell) \tilde N_i(\ell)$. We will apply this double-copy construction to the 
two-loop five-point amplitudes in ten-dimensional SYM and the resulting type II supergravities, see
section~\supgrav. When dressed with non-supersymmetric numerators $\tilde N_i(\ell)$ of pure Yang--Mills,
our results for $N_i(\ell)$ might serve as a convenient starting point to investigate two-loop five-point
amplitudes in half-maximal supergravity.

\subsec Ten-dimensional SYM

\noindent
Ten-dimensional SYM can be described in a super-Poincar\'e covariant manner using  a
set of superfields $\{A_\a, A^m, W^\a, F_{mn}\}$ \refSYM. They depend on the ten-dimensional
superspace coordinates $\{x^m,\theta^\a\}$ with vector and spinor indices $m,n=0,1,\ldots,9$
and $\alpha,\beta=1,2,\ldots,16$ of the Lorentz group. The linearized equations of
motion \refSYM\ in momentum space\foot{Our conventions for (anti-)symmetrizing $n$ vector indices include a factor of $1/n!$.}
\eqnn\SYM
$$\displaylines{
\hfill D_\a A_\b + D_\b A_\a = \g^m_{\a\b} A_m, \qquad D_\a A_m = (\g_m W)_\a + k_m A_\a  \hfill\phantom{(1.1)}\cr
\hfill D_\a F_{mn} = 2k_{[m} (\g_{n]} W)_\a, \qquad  D_\a W^{\b} = {1\over 4}(\g^{mn})_\a{}^\b F_{mn}  \hfill\SYM\cr
}$$
involve light-like momenta $k^2 =0$ via plane waves $e^{k\cdot x}$ and fermionic covariant derivatives
\eqn\coder{
D_{\alpha} \equiv { \partial \over \partial \theta^\alpha} + {1\over 2} (\gamma^m \theta)_\alpha \ , \ \ \ \ \ \ \{ D_\alpha , D_\beta \} = \gamma^m_{\alpha\beta} \partial_m  \ ,
}
with $16 \times 16$ Pauli matrices\foot{Pauli matrices frequently appear in their
antisymmetrized combinations with symmetry properties $\gamma^{mnp}_{\alpha \beta}=-\gamma^{mnp}_{\beta \alpha }$
and $\gamma^{mnpqr}_{\alpha \beta}=\gamma^{mnpqr}_{\beta \alpha }$ which are normalized
like $\gamma^{mn}{}_{\a}{}^{\b} \equiv {1\over 2} (\gamma^m\gamma^n - \gamma^n \gamma^m)_{\a}^{\b}$.}
$\gamma^m_{\alpha \beta}=\gamma^m_{\beta \alpha }$ subject to the Clifford algebra $\gamma^{(m}_{\alpha \beta}\gamma^{n)\beta \gamma} = \eta^{mn} \delta^\gamma_\alpha$.

The equations of motion \SYM\ identify
$8+8$ on-shell degrees of freedom associated with a gluon and gluino. Their polarizations
are described by a transverse vector $k_me^m=0$ and a spinorial solution to the massless
Dirac-equation $k_m \gamma^m_{\alpha \beta} \chi^\beta =0$, respectively. In a gauge
where $\theta^\alpha A_\alpha=0$, they enter the explicit $\theta$-expansions via \thetaSYM
\eqn\expl{
A_\alpha= e^{k \cdot x} \Big({1\over 2}e_m(\g^m\t)_\a -{1\over 3}(\chi \g_m\t)(\g^m\t)_\a
-{1\over 16}k_{m} e_{n}(\g_p\t)_\a (\t\g^{mnp}\t) + \cdots \Big)  \ .
}

\subsec Multiparticle superfields

Ten-dimensional SYM arises from the massless sector of the open pure spinor superstring \superpoincare.
Its $i^{\rm th}$ scattering state is represented by the integrated vertex operator $U_i$ which
involves all of $\{A^i_\a, A_i^m, W_i^\a, F^i_{mn}\}$ along with various worldsheet fields. In string
calculations, the latter act on additional vertex operators via operator product expansions (OPEs) and
build up so-called multiparticle superfields \EOMBBs
\eqnn\twopart
$$\eqalignno{
A_{12}^\alpha&\equiv - \half\bigl[ A^1_\a (k^1\cdot A^2) + A^1_m (\g^m W^2)_\a - (1\leftrightarrow 2)\bigr] \cr
A^{12}_m&\equiv\half\Bigl[ A^1_p F^2_{pm} - A^1_m(k^1\cdot A^2) + (W^1\g_m W^2) - (1\leftrightarrow 2)\Bigr]
\cr
W_{12}^\alpha &\equiv {1\over 4}(\g^{mn}W^2)^\a F^1_{mn} + W_2^\a (k^2\cdot A^1) - (1\leftrightarrow 2)
&\twopart \cr
F_{12}^{mn} &\equiv F_2^{mn}(k^2\cdot A^1) +  F_2^{[m}{}_{p}F^{n]p}_1 + k_{12}^{[m}(W_1\g^{n]}W_2) -
(1\leftrightarrow 2) \ .
}$$
In the point-particle limit, these string-inspired generalizations of linearized
SYM accompany the tree-level subdiagrams seen in \figone. They capture, for
instance, the reducible parts of the subsequent five-point two-loop amplitudes
and have generalizations $\{A^{12\ldots p}_\a, A_{12\ldots p}^m, W_{12\ldots p}^\a, F^{12\ldots p}_{mn}\}$
to any number of states \EOMBBs, see \figone. We will use multiparticle
labels such as $B=12\ldots p$ to keep the multiplicity of superfields unspecified,
i.e. $W^\alpha_B$ can become either $W_1^\alpha$ or $W_{12}^\alpha$ upon setting $B=1$ or $B=12$. 


\ifig\figone{Interpretation of multiparticle superfields as tree-level subdiagrams where the external states are represented by multiparticle labels $B=12\ldots p$.}
{\epsfxsize=0.55\hsize\epsfbox{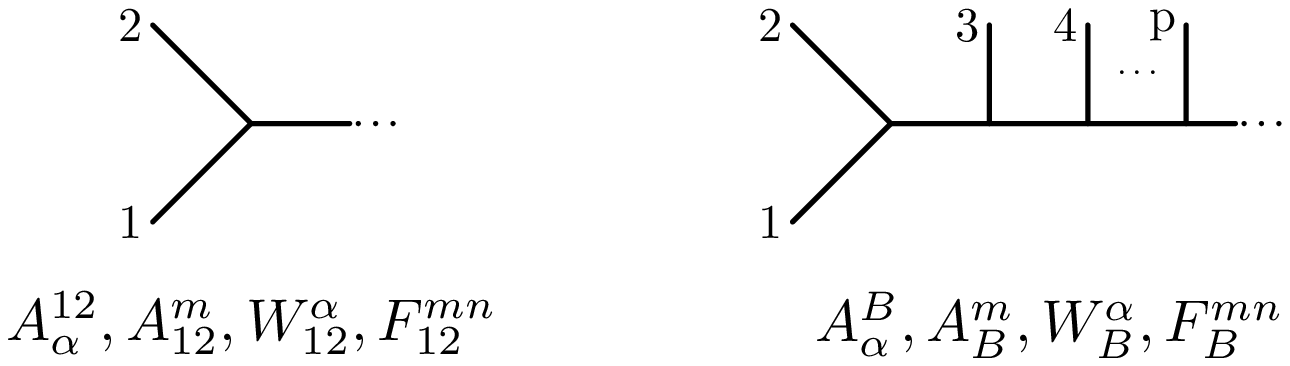}}

The non-linear version of the equations of motion \SYM\ are solved by the generating
series of multiparticle fields \HighSYM. The non-linearities are reflected by the
contact terms $\sim (k^1\cdot k^2)$ in their multiparticle equations of motion
\eqnn\EOMAmtwo
$$\eqalignno{
2 D_{(\a} A^{12}_{\b)} &= \g^m_{\a\b}A^{12}_m + (k^1\cdot k^2)(A^1_\a A^2_\b + A^1_\b A^2_\a)  \cr
D_\a A^{12}_m &= (\g_m W^{12})_\a + k^{12}_m A^{12}_\a + (k^1\cdot k^2)(A^1_\a A^2_m - A^2_\a A^1_m) \cr
D_\a W^\b_{12} &= {1\over 4}(\g^{mn})_\a{}^\b F^{12}_{mn} + (k^1\cdot k^2)(A^1_\a W_2^\b - A^2_\a W^\b_1) &\EOMAmtwo\cr
D_\a F^{12}_{mn} &= k^{12}_m (\g_n W^{12})_\a - k^{12}_n (\g_m W^{12})_\a + (k^1\cdot k^2)(A^1_\a F^2_{mn} - A^2_\a
F^1_{mn})  \cr
& \ \  + (k^1\cdot k^2)( A^{1}_{n} (\g_{m} W^2)_\a - A^{2}_{n} (\g_{m} W^1)_\a - A^{1}_{m} (\g_{n} W^2)_\a + A^{2}_{m} (\g_{n} W^1)_\a) \ ,
}$$
where the notation $k_{12\ldots p}^m \equiv k_1^m + k_2^m +\ldots+k_p^m$ will be
used throughout this work.

\subsec Pure spinor superspace

The physical components of superspace expressions in SYM can be conveniently
extracted using a bosonic pure spinor $\lambda^\alpha$ which is defined to obey \bigHowe
\eqn\PScon{
(\lambda\gamma^m \lambda) = 0 \ .
}
Corollaries of this pure spinor constraint include
$(\lambda\gamma^m)_\alpha (\lambda \gamma_m)_\beta=0=(\lambda\gamma^m \gamma^{pq}\lambda) $
which will frequently enter the subsequent manipulations. Pure spinor superspace is
defined by an extension of the standard ten-dimensional superspace coordinates $\{x^m, \t^\a\}$ to $\{x^m ,\theta^\alpha,\lambda^\alpha\}$
subject to the following component prescription \superpoincare
\eqn\prescr{
\langle  (\lambda \gamma^m \theta)
 (\lambda \gamma^n \theta)
  (\lambda \gamma^p \theta)
   (\lambda \gamma_{mnp} \theta) \rangle = 2880 \ .
}
The order $\lambda^3 \theta^5$ is singled out by supersymmetry and the cohomology of the BRST operator
\eqn\qcharge{
Q \equiv \lambda^\alpha D_\alpha \ .
}
BRST-closed superfields $QS(\theta,\lambda)=0$ give rise to supersymmetric and gauge
invariant components $\langle S(\theta,\lambda) \rangle$ under \prescr\ whereas
BRST-exact superfields $E(\theta,\lambda) = Q\Sigma(\t,\l)$ yield $\langle E(\theta,\lambda) \rangle =0$ \superpoincare.
Expressions of order $\lambda^3 \theta^5$ with a different tensor structure as compared
to \prescr\ are uniquely fixed by group theory since this tensor product contains
only one Lorentz scalar. The required group theory manipulations are automated in \PSS.

As the central idea of this work and preceding papers \refs{\towards, \MafraJQ, \MafraGJA}, scattering amplitudes
$\langle S(\theta,\lambda) \rangle$ in ten-dimensional SYM are proposed by
constructing a BRST-invariant superfield $S(\theta,\lambda)$ whose kinematic
poles reproduce the expected Feynman diagrams. In this approach, the superfield 
$S(\theta,\lambda)$ carries the kinematical data of state $i$ through the
superfields $\{A^i_\a, A_i^m, W_i^\a, F^i_{mn}\}$ whose equations of motion
\SYM\ determine the BRST variation.

For example, the unintegrated vertex operator of the superstring,
\eqn\Vone{
V_i \equiv \lambda^\alpha A^i_\alpha \ , \ \ \ \ \ \ Q V_i = 0 \ ,
}
suffices to write down the three-point tree-level subamplitude in pure spinor superspace,
\eqn\tree{
A^{{\rm tree}}(1,2,3)=\langle V_1 V_2 V_3 \rangle = (e_1\cdot e_2) (e_3 \cdot k_1)
+ e_1^m(\chi_2 \gamma_m \chi_3) + {\rm cyc}(1,2,3) \ .
}
This sample calculation based on the $\theta$-expansion \expl\ and the
prescription \prescr\ illustrates how all the component amplitudes are
supersymmetrically embedded into BRST-closed superfields such as $V_1 V_2 V_3$.
We will limit our subsequent discussion of two-loop amplitudes to their superspace representatives since the component extraction for any superfield
numerator can be performed in an automated way \refs{\PSS,\FORM}, and the resulting components can be downloaded from \WWW.

\newsec{Field-theory amplitudes and BRST cohomology at two-loops}
\par\seclab\sectwo

\noindent Multiloop superstring amplitudes computed with the pure spinor formalism give rise to superspace expressions in the cohomology of the pure spinor BRST charge \refs{\multiloop,\NMPS}. As explained in \superpoincare, superfields in the BRST 
cohomology translate to gauge invariant and supersymmetric component expansions. Since 
ten-dimensional SYM and type II supergravities arise in a certain limit of superstring 
theories, their scattering amplitudes belong to the BRST cohomology as well. Multiloop 
integrands for these field-theory amplitudes are strongly
constrained by demanding BRST invariance and the propagator structure expected from Feynman diagrams.
Together with a string-inspired set of admissible kinematic building blocks, these requirements will
allow us to fix the two-loop five-point amplitudes.

\subsec BRST properties of kinematic numerators

Inspired by the discussion of section \secBCJ, multiloop amplitudes of
ten-dimensional SYM theory are organized in terms of cubic graphs
\eqn\cubicBagain{
{\cal A}^{g-{\rm loop}}_n = \int \prod_{j=1}^g {d^D \ell_j \over (2\pi)^D}
\sum_{\Gamma_i} {\langle N_i (\ell)\rangle C_i \over \prod_k P_{k,i}(\ell)} \ .
}
Maximal supersymmetry suppresses any graph $\Gamma_i$ with a triangle, bubble or tadpole subdiagram
\BernZX. In the superspace setup of this paper, the numerators $N_i(\ell)$ will be given by {\it local}
pure spinor superspace expressions, whose form is suggested by the propagator structure $P_{k,i}(\ell)$
of its corresponding cubic graph. Requiring BRST invariance of the
integrand in \cubicBagain\ largely determines the mapping between cubic graphs and
superspace numerators, and the subsequent examples are completely fixed when assuming 
a string-inspired ansatz for admissible kinematic building blocks.

If individual numerators are not BRST invariant by themselves, their BRST
variations must lead to cancellations among different graphs
to yield an overall BRST-invariant 
integrand. That is only possible if the BRST variation $QN_i(\ell)$ cancels one of its propagators
$P_{k,i}(\ell)$. Therefore, the superspace expression of $N_i(\ell)$ is constrained by the following
requirement:
\eqn\principle{
\hbox{\it each term of $QN_i(\ell)$ must contain a factor of $P_{k,i}(\ell)$.}
}
Otherwise the BRST variation of the integrand \cubicBagain\ would have a non-vanishing residue at the
simultaneous pole $\prod_k P_{k,i}(\ell)$ and could not vanish. 

We will use the following notation to distinguish between superspace integrands and integrated expressions,
\eqnn\defintegrand
$$\eqalignno{
{\cal A}^{2-{\rm loop}}_n &= \int  {d^D\ell \ d^Dr \over (2\pi)^{2D}} \langle {\cal A}^{2-{\rm loop}}_n(\ell,r) \rangle &\defintegrand
\cr
 A^{{\rm 2-loop}}(1,2,3, \ldots,n) &= \int {d^D\ell \ d^Dr \over (2\pi)^{2D}} \langle A^{{\rm 2-loop}}(1,2,3, \ldots,n| \ell,r) \rangle   \ ,
}$$
where $A^{{\rm 2-loop}}(1,2,3,
\ldots,n)$ denotes the planar single-trace contribution of ${\cal A}^{2-{\rm loop}}_n$ in the trace
basis of $C_i$ \dixon. In the following sections we will use the method outlined above to construct the
SYM and supergravity five-point two-loop integrands. They lead to BRST-closed integrated amplitudes
once the freedom to shift or rename the integration variables $\ell, r$ is taken into
account\foot{The diagrammatic bookkeeping of BRST variations in the appendix \appBRST\ automatically freezes this
freedom by means of the automorphism symmetries of the different integrand topologies after the
cancellation of propagators resulting from the BRST variation of the numerators.}.

\subsec The SYM two-loop four-point amplitude

 
 \ifig\figtwofour{The planar integrand of the SYM two-loop four-point amplitude.}
{\epsfxsize=0.75\hsize\epsfbox{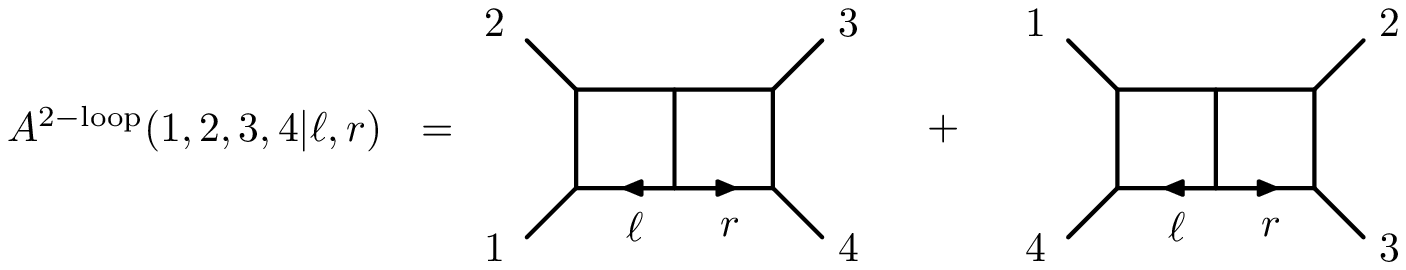}}

\noindent Recall that the two-loop four-point amplitude of the pure spinor superstring
\refs{\twoloop,\twolooptwo,\coefftwo} can
be written in terms of a single kinematic factor
\eqnn\twoloBT
$$\eqalignno{
T_{1,2|3,4} &\equiv {1\over 64}
(\lambda \gamma_{mnpqr} \lambda) F^{mn}_1 F^{pq}_2 \big[ F^{rs}_3(\lambda \gamma_s W_4)+
F^{rs}_4(\lambda \gamma_s W_3)\big] + (1,2\leftrightarrow 3,4) \cr
QT_{1,2|3,4} &=0 \ ,&\twoloBT
}$$
whose BRST invariance follows from the equations of motion \SYM\ and the pure spinor constraint \PScon.
Since superstring amplitude reduce to SYM amplitudes in field-theory limit, the most
natural expression for the planar four-point two-loop integrand is given by
\eqnn\deli
$$\eqalignno{
A^{2-{\rm loop}}(1,2,3,4|\ell,r)&=
{ T_{1,2|3,4}\over \ell^2 r^2 (\ell+r)^2 \, (\ell-k_1)^2 (\ell-k_{12})^2 \, (r-k_4)^2 (r-k_{34})^2} \cr
&+ { T_{4,1|2,3}\over \ell^2 r^2 (\ell+r)^2 \, (\ell-k_4)^2
(\ell-k_{41})^2 \, (r-k_3)^2 (r-k_{23})^2}\,.
&\deli
}$$
The contributing double-box graphs are depicted in \figtwofour, and BRST invariance of its
numerators is inherited from \twoloBT. Furthermore, it has been shown at the superspace level that \mafraids
\eqn\twoloF{
 \langle T_{1,2|3,4} \rangle = s_{12}^2 s_{23} A^{{\rm tree}}(1,2,3,4) \ ,
 }
hence, the amplitude \deli\ agrees with the result of \BernTwoLoop. The Mandelstam invariants are
\eqn\mandconv{
s_{ij} \equiv (k_i\cdot k_j) = \half (k_i+k_j)^2 \ , \ \ \ \ s_{i_1 i_2\ldots i_p }
\equiv \half (k_{i_1}+k_{i_2}+ \ldots + k_{i_p})^2 \,.
}
A rigorous derivation of \deli\ as the field-theory limit of the two-loop open superstring
amplitude \twoloop\ is expected to closely follow the closed-string discussion in \tropical.

\subsec Five-point building blocks in pure spinor superspace

By the aforementioned no-triangle property of maximal SYM \BernZX, the two-loop five-point 
amplitude involves double-box and penta-box diagrams along with their non-planar 
counterparts. The numerators for the pentagon subdiagrams are known to depend linearly 
on the loop momentum \CarrascoMN, and therefore the five-point amplitude will require 
vector building blocks in addition to scalar building blocks.

For the scalar building block, one can use the multiparticle version of \twoloBT
\eqn\twoloB{
T_{A,B|C,D} \equiv {1\over 64}
(\lambda \gamma_{mnpqr} \lambda) F^{mn}_A F^{pq}_B \big[ F^{rs}_C(\lambda \gamma_s W_D)
+ F^{rs}_D(\lambda \gamma_s W_C)\big] + (A,B\leftrightarrow C,D) \ .
}
The symmetry properties of $T_{1,2|3,4}$ described in \twoloop\ do not depend on the single-particle
nature of the superfields and directly carry over to \twoloB,
\eqn\twoloC{
T_{A,B|C,D}= T_{B,A|C,D}=T_{C,D|A,B} \ , \quad T_{A,B|C,D}+T_{B,C|A,D}+T_{C,A|B,D}=0 \,.
}
The latter follows from the gamma-matrix manipulation
$(\lambda \gamma_{[mnpqr} \lambda) (\lambda \gamma_{s]})_\alpha = 0$. For five
points, the BRST variation of \twoloB\ follows from multiparticle equations of motion \twopart,
\eqn\twoloE{
Q T_{12,3|4,5} = s_{12} ( V_1 T_{2,3|4,5} - V_2 T_{1,3|4,5} ) \ ,
}
and the terms without a factor of $s_{12}$ drop out by virtue of the pure spinor constraint.

In analogy to the one-loop vector building block of \FiveSdual, the scalar two-loop building blocks
\twoloB\ allow for a vector counterpart\foot{The following definition can be generalized to
multiparticle level $T_{1,2,3|4,5}^m \rightarrow T_{A,B,C|D,E}^m$ if the corresponding uplift of \mtwod\
is projected to the symmetries of \delm.} suitable to represent linear dependencies on $\ell$,
\eqn\twoloH{
T^m_{1,2,3|4,5}\equiv   A_1^m T_{2,3|4,5} + A_2^m T_{1,3|4,5} + A_3^m T_{1,2|4,5} + 
W^m_{1,2,3|4,5}  \ .
}
The last summand $W^m_{1,2,3|4,5}$ is designed to cancel the term $(\lambda \gamma^m W_1)$ within $QA^m_1$, i.e. 
\eqnn\dell
$$\eqalignno{
QW^m_{1,2,3|4,5}&=   - (\lambda \gamma^m W_1) T_{2,3|4,5}  - (\lambda \gamma^m W_2) T_{1,3|4,5}
- (\lambda \gamma^m W_3) T_{1,2|4,5}  \ .
&\dell
}$$
In a symmetrization convention where $W^\alpha_{(1}F^{mn}_{2)}\equiv W^\alpha_{1}F^{mn}_{2}+W^\alpha_{2}F^{mn}_{1}$,
its explicit superfield representation is given by
\eqnn\mtwod
$$\eqalignno{
W_{3,4,5|1,2}^m &\equiv
{1\over 48}   (\lambda \gamma_{pq} \gamma^m W_{(1}) F_{2)}^{pq}(\lambda \gamma_r W_5)(\lambda \gamma_s W_{(3}) F_{4)}^{rs}\cr
&- {1\over 128} (\lambda \gamma^m W_5) (\lambda \gamma_{pq}
\gamma^r W_{(1}) F^{pq}_{2)}
(\lambda \gamma_{st} \gamma_r W_{(3}) F^{st}_{4)} &\mtwod\cr
&+{1\over 96}(W_3 \gamma^{mst} W_4) (\lambda \gamma_{npqrs} \lambda)(\lambda \gamma_t W_5) F_1^{np} F_2^{qr}  + (5\leftrightarrow 3,4) \ .
}$$
The BRST variation of $T^m_{1,2,3|4,5}$ in \twoloH\ then connects with the scalar counterpart \twoloE,
\eqn\QQTm{
Q T^m_{1,2,3|4,5} = k_1^m V_1 T_{2,3|4,5} + k_2^m V_2 T_{1,3|4,5} + k_3^m V_3 T_{1,2|4,5} \ .
}
The symmetries \twoloC\ of the scalar building block and the form
of $W^m_{1,2,3|4,5}$ in \mtwod\ imply
\eqn\delm{
T^m_{1,2,3|4,5}=T^m_{(1,2,3)|(4,5)}\,,\quad
\langle T^m_{1,2,3|4,5} \rangle = \langle T^m_{3,4,5|1,2}+T^m_{2,4,5|1,3}+T^m_{1,4,5|2,3} \rangle \ .
}
Note that the two-loop building blocks of this section are functionals of all the external labels, 
e.g. $T^m_{i,j,k|p,q}$ can be specialized to any permutation $(i,j,k,p,q)$ of 
$(1,2,3,4,5)$. This surpasses the constraints on superspace numerators at tree-level 
\refs{\towards, \MafraJQ} and one-loop \MafraGJA\ where individual external legs need to 
be globally associated with unintegrated vertices \Vone\ or their multiparticle versions \EOMBBs.

\newsec{The two-loop five-point amplitudes in SYM and supergravity}

In this section, we assemble the five-point two-loop amplitudes from the six topologies of cubic
diagrams without triangles \CarrascoMN\ depicted in \figoverview.

\ifig\figoverview{The six topologies of two-loop five-point diagrams whose 
color factors are given in \allcolor. In a BCJ representation subject to \BCJform, the numerators satisfy
$N^{(b)}=-N^{(a)}$ as well as $N^{(f)}=N^{(e)} = -N^{(d)}$, and the explicit superspace expressions for $N^{(d)}$
and $N^{(a)}$ are given in \Nddef\ and \Nadef, respectively.}
{\epsfxsize=0.80\hsize\epsfbox{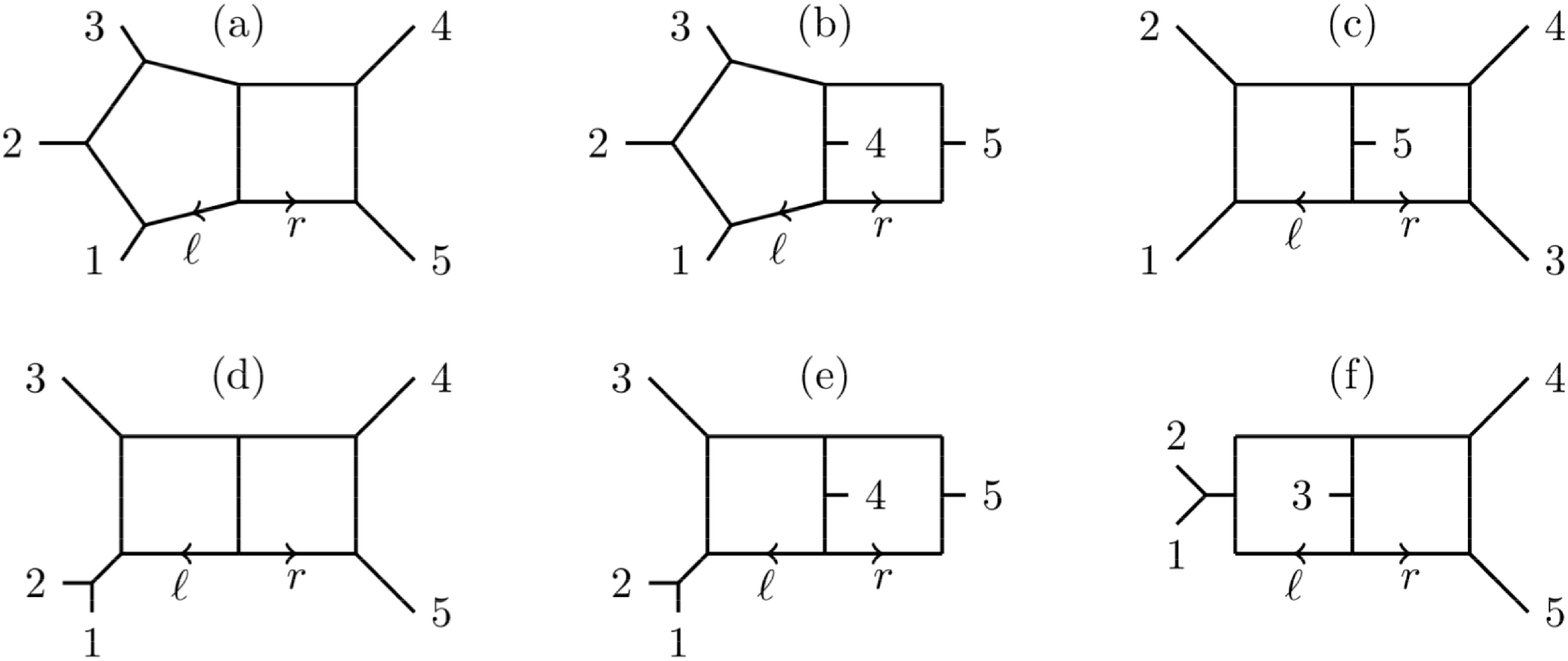}}

\subsec Color factors

The color factors $C^{(a)}_{12345},\ldots,C^{(f)}_{12345}$ associated to the diagrams of \figoverview\
encompass seven factors of $f^{abc}$, the structure constants of the gauge group. Their contractions
follow from dressing each cubic vertex\foot{The color factors $C^{(b)}_{12345}$, $C^{(e)}_{12345}$ and
$C^{(f)}_{12345}$ contain an extra minus sign w.r.t. \CarrascoMN\ due to our choice of drawing graphs
without crossing lines.} of the diagram with a factor of $f^{abc}$ and each internal line with a
Kronecker delta in the adjoint representation of the gauge group,
\eqn\allcolor{
\eqalign{
C^{(a)}_{12345}&=f^{4 e c} f^{5 b e} f^{a 1 h} f^{b a d} f^{c d g} f^{g j 3} f^{h 2 j} \,,\cr
C^{(c)}_{12345}&=f^{1 a c} f^{2 h c} f^{a d g} f^{b 4 j} f^{e d 5} f^{g j 3} f^{h b e} \,,\cr
C^{(e)}_{12345}&=f^{4 e c} f^{5 d b} f^{a j h} f^{b e a} f^{c d g} f^{g j 3} f^{h 1 2}\,,
}\quad\eqalign{
C^{(b)}_{12345}&= f^{4 d e} f^{b 5 c} f^{a 1 h} f^{c d g} f^{e b a} f^{g j 3} f^{h 2 j}\,,\cr
C^{(d)}_{12345}&= f^{4 e c} f^{5 b e} f^{a j h} f^{b a d} f^{c d g} f^{g j 3} f^{h 2 1}\,,\cr
C^{(f)}_{12345}&= f^{1 2 c} f^{a d b} f^{b j 5} f^{c a g} f^{e 3 d} f^{g h e} f^{h 4 j}\,.
}}
Using the procedure described in \dixon\ they can be translated into a trace basis leading to color structures of
the form $N_c^2 {\rm Tr}(t^1t^2t^3t^4t^5)$,
${\rm Tr}(t^1t^2t^3t^4t^5)$, $N_c {\rm Tr}(t^4t^5) {\rm Tr}(t^1t^2t^3)$, ${\rm Tr}(t^4) {\rm Tr}(t^5) {\rm Tr}(t^1t^2t^3)$ and $N_c {\rm Tr}(t^5) {\rm Tr}(t^1t^2t^3t^4)$. The gauge group is left completely general at this point such that its generators $t^i$ are not necessarily traceless. The number of colors $N_c$ stems from the trace of the identity matrix.

\ifig\generaldb{The numerator for the general massive double-box is
given by $T_{A,B|C,D}$ in \twoloB.}
{\epsfxsize=0.60\hsize\epsfbox{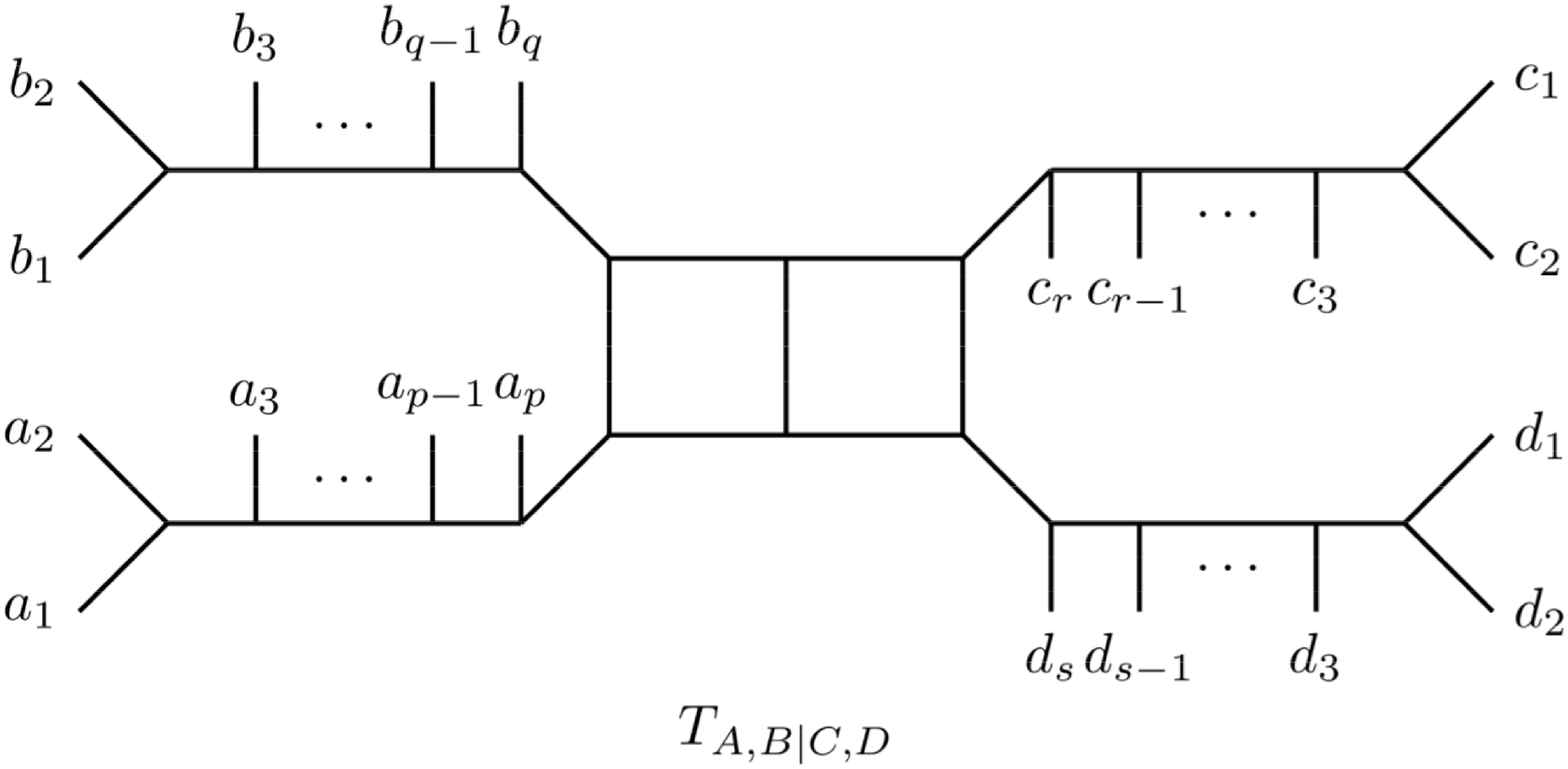}}

\subsec Planar cubic graphs and superspace numerators

The BRST cohomology principle \principle\ suggests the scalar multiparticle building block $T_{A,B|C,D}$ in
\twoloB\ to furnish a massive double-box numerator, see \generaldb.
This conjectural identification is based on the fact that all the $\ell$-independent propagators in \generaldb\ are canceled by some summand of $Q T_{A,B|C,D}$, see e.g. \twoloE. Moreover, the symmetry
$T_{A,B|C,D}=T_{A,B|D,C}$ is compatible with the no-triangle property and the kinematic Jacobi identity:

\centerline{{\epsfxsize=0.60\hsize\epsfbox{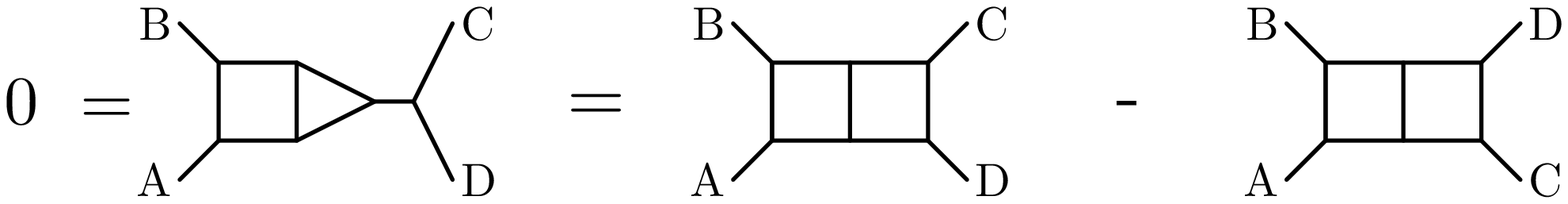}}}
\smallskip

\noindent The dictionary in \generaldb\ implies that the planar five-point double-box in {\figoverview}(d) is represented by the numerator
\eqn\Nddef{
N^{(d)}_{12,3|4,5}\equiv T_{12,3|4,5}  \,.
}
The scalar and vector BRST transformations in \twoloE\ and \QQTm\ allow to construct a candidate for the penta-box numerator
of {\figoverview}(a) as
\eqnn\Nadef
$$\eqalignno{
N^{(a)}_{1,2,3|4,5}(\ell) &\equiv {1\over 2}(\ell_m + \ell_m - k_m^{123}) T^m_{1,2,3|4,5}  
+ {1\over 2} (T_{12,3|4,5} + T_{13,2|4,5}+T_{23,1|4,5}) \ .
&\Nadef
}$$
This expression is designed from the cohomology principle \principle\ since $\ell$-dependent
propagators are canceled under the BRST variation,
\eqnn\delr
$$\eqalignno{
QN^{(a)}_{1,2,3|4,5}(\ell) &=  {1\over 2} \Big( V_1 T_{2,3|4,5} \big[ \ell^2 - (\ell-k_{1})^2 \big]
+V_2 T_{1,3|4,5} \big[ (\ell-k_{1})^2 - (\ell-k_{12})^2 \big] \cr
&\ \ \ \ \ \ \ \ \  +V_3 T_{1,2|4,5} \big[ (\ell-k_{12})^2 - (\ell-k_{123})^2 \big] \Big) \,.
&\delr
}$$
Both the composition \Nadef\ from scalar and vector superfields and the form of the BRST
variation \delr\ resemble the one-loop pentagon numerator, see (4.5) and (4.6) of \MafraGJA.

As illustrated in \figDbox, the numerator in \Nadef\ depends on the averaged loop momentum $\ell_m +
(\ell_m - k_m^{123})$ from the two terminal edges of the pentagonic worldline segment. This makes sure
that the numerator inherits the reflection symmetry of the cubic diagram in {\figoverview}(a)
\CarrascoMN,
\eqn\symmPB{
N^{(a)}_{1,2,3|4,5}(\ell) = - N^{(a)}_{3,2,1|5,4}(k_{123}-\ell) \ . 
}
Also, the symmetry of $N^{(a)}_{1,2,3|4,5}(\ell)$ in $4,5$ is compatible with the no-triangle property:
\medskip
\centerline{{\epsfxsize=0.65\hsize\epsfbox{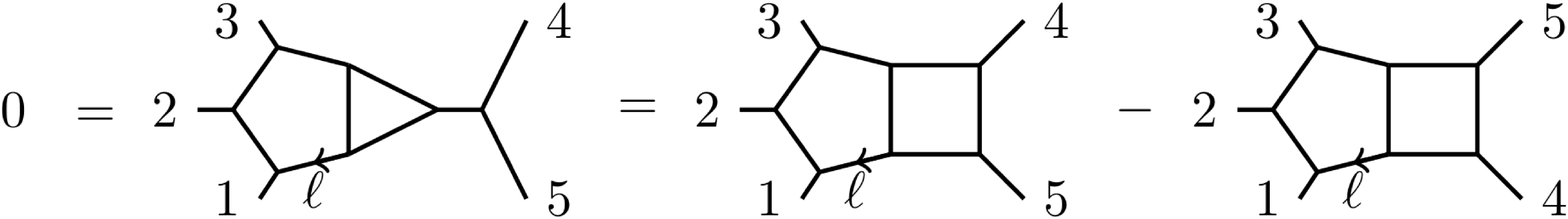}}}
\smallskip

\noindent Finally it is easy to show the kinematic Jacobi identity \CarrascoMN
\eqn\JacDbox{
N^{(d)}_{12,3|4,5} = N^{(a)}_{1,2,3|4,5}(\ell) - N^{(a)}_{2,1,3|4,5}(\ell) \ ,
}
which interlocks antisymmetrized penta-boxes with double-boxes.

\subsec Planar two-loop amplitudes in SYM

\noindent
The numerators \twoloB\ and \Nadef\ for planar double-box and penta-box diagrams are 
summarized in \figDbox\ and satisfy the BRST cohomology principle \principle. Hence, we 
propose the following single-trace five-point integrand for \defintegrand:
\eqnn\dels
$$\eqalignno{
A^{2-{\rm loop}}(1,2,3,4,5|\ell,r) &=
{1\over \ell^2 r^2 (\ell+r)^2}\Big[ { N^{(a)}_{1,2,3|4,5}(\ell) \over  (\ell-k_1)^2 (\ell-k_{12})^2 (\ell-k_{123})^2 \, (r-k_5)^2 (r-k_{45})^2} \cr
&+  { N^{(d)}_{12,3|4,5} \over k_{12}^2 (\ell-k_{12})^2 (\ell-k_{123})^2 \, (r-k_5)^2
(r-k_{45})^2 } &\dels\cr
& +  {N^{(d)}_{5,12|3,4} \over  k_{12}^2 (\ell-k_{5})^2 (\ell-k_{512})^2 \, (r-k_4)^2
(r-k_{34})^2} + {\rm cyc}(1,2,3,4,5)\Big]
  \ .
}$$
Given that ${\rm cyc}(1,2,3,4,5)$ instructs to add the four cyclic images of $(1,2,3,4,5)$, the expression
in \dels\ is manifestly cyclic, and its BRST invariance
is easily checked with a diagrammatic
bookkeeping of the associated integrals, see the appendix \appBRST\ for more details.

\ifig\figDbox{The mapping between the double-box and penta-box graphs and superspace numerators.
The vertical bar in the notation $\ldots|\ldots$ separates the two worldline segments.}
{\epsfxsize=0.57\hsize\epsfbox{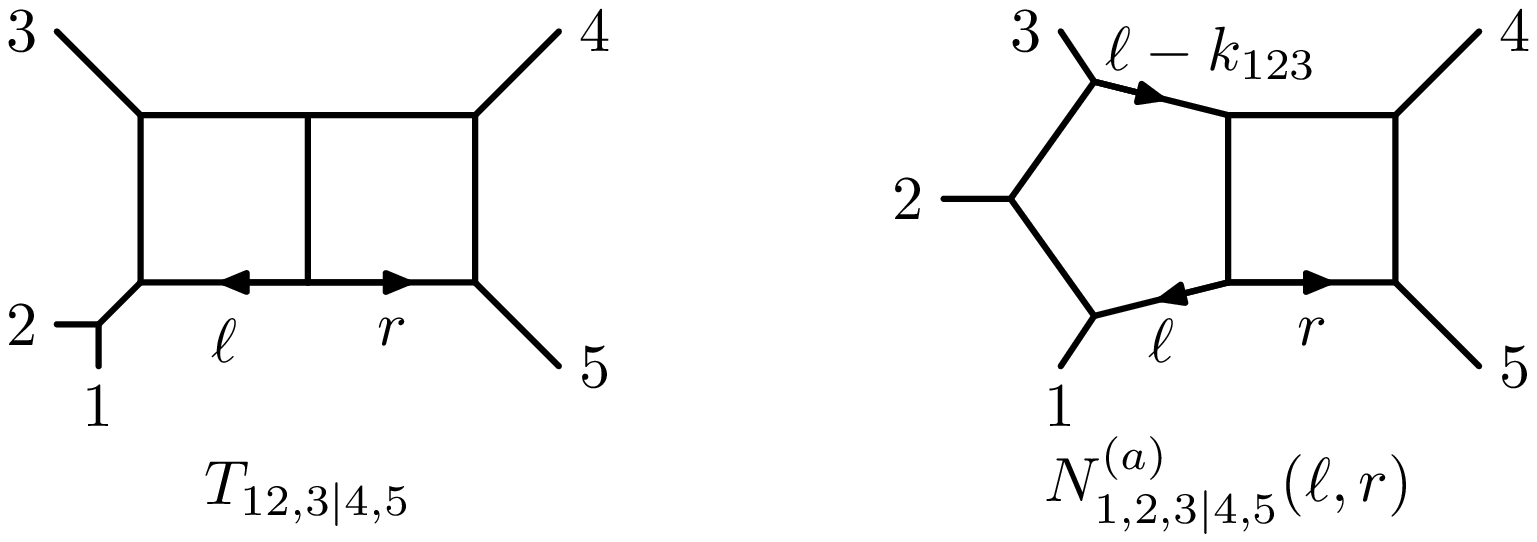}}

\subsec Non-planar cubic graphs and superspace numerators

\noindent
In contrast with the one-loop level \BernZX, single-trace subamplitudes at two-loops are no longer
sufficient to infer the kinematic structure associated with all color tensors. Candidate expressions
for the non-planar diagrams in term of the planar numerators can be obtained from the BCJ duality and have been worked out for the
five-point two-loop case in \CarrascoMN. Non-planar double-boxes follow from the no-triangle property and the kinematic Jacobi
relation shown in \fignpldouble,
\eqn\Nef{
N^{(e)}_{12,3|4,5} = N^{(f)}_{12,3|4,5} = - N^{(d)}_{12,3|4,5}\,.
}
\ifig\fignpldouble{The derivation of $N^{(e)}_{12,3|4,5} = N^{(f)}_{12,3|4,5} = - N^{(d)}_{12,3|4,5}$ from a
kinematic Jacobi identity and the no-triangle property. A similar analysis leads
to $N^{(b)}_{1,2,3|4,5} = - N^{(a)}_{1,2,3|4,5}$. }
{\epsfxsize=0.7\hsize\epsfbox{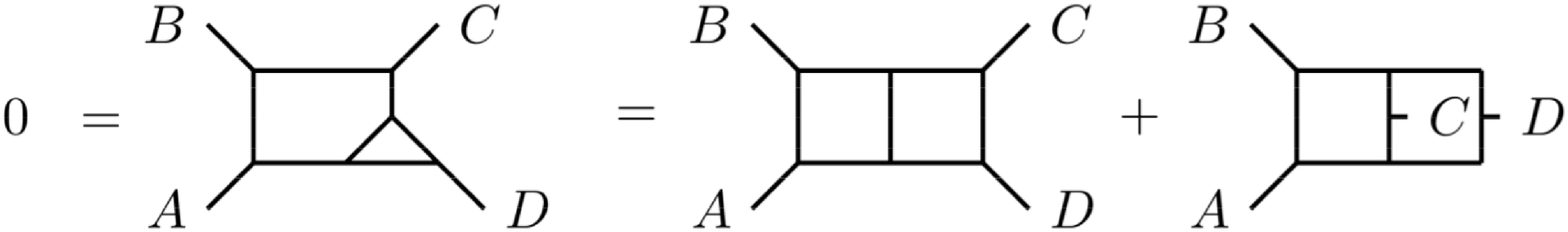}}


\noindent The numerators for non-planar penta-box diagrams are obtained from a sequence
of similar BCJ moves and given by \CarrascoMN
 \eqnn\nplazero
  \eqnn\Ncdef
$$\eqalignno{
N^{(b)}_{1,2,3|4,5}(\ell)&= -N^{(a)}_{1,2,3|4,5}(\ell) &\nplazero \cr
N^{(c)}_{1,2|4,3|5}(\ell,r)&= N^{(a)}_{1,2,5|3,4}(\ell) + N^{(a)}_{3,4,5|1,2}(r)&\Ncdef
 \cr
&={1\over 2}(\ell_m + \ell_m - k_m^{125})T^m_{1,2,5|3,4}+{1\over 2}(r_m+r_m - k_m^{345})T^m_{3,4,5|1,2}\cr
&\quad{}+{1\over 2}(T_{12,5|3,4}+T_{25,1|3,4}+T_{15,2|3,4}+T_{34,5|1,2}+T_{35,4|1,2}+T_{45,3|1,2})\,.
}$$
As a non-trivial consistency condition on the superspace numerator \Ncdef, one can check 
its compatibility with the BRST principle \principle\ through the BRST variation
\eqnn\nplb
$$\eqalignno{
2QN^{(c)}_{1,2|4,3|5}(\ell,r)&=
V_5 T_{1,2|3,4} \big[(\ell+r+k_5)^2 - (\ell+r)^2 \big] 
&\nplb
 \cr
&+V_1 T_{2,5|3,4} \big[\ell^2 - (\ell-k_1)^2 \big] +V_2 T_{1,5|3,4} \big[(\ell-k_1)^2 - (\ell-k_{12})^2 \big] 
 \cr
&+V_3 T_{4,5|1,2} \big[r^2 - (r-k_3)^2 \big] +V_4 T_{3,5|1,2} \big[(r-k_3)^2 - (r-k_{34})^2 \big]  \ .
}$$
Moreover, \Ncdef\ satisfies the required self-symmetries \CarrascoMN. For example, a rotation by
180$^{\circ}$ (with a simultaneous flip of the vertex next to particle 5) maps the diagram to itself up to relabellings:
\eqnn\nplc
$$\eqalignno{
N^{(c)}_{1,2|4,3|5}(\ell,r)&=- N^{(c)}_{4,3|1,2|5}(k_{34}-r,k_{12}-\ell)
&\nplc
}$$
This is respected by the expression in \Ncdef\ provided that\foot{Surprisingly, \npld\ is also the
condition for BRST invariance of the corresponding closed string five-point amplitude at two loops.}
\eqnn\npld
$$\eqalignno{
k_m^5 \langle T^m_{3,4,5|1,2}+T^m_{1,2,5|3,4}\rangle&= \langle T_{15,2|3,4}+T_{25,1|3,4}+T_{35,4|1,2}+T_{45,3|1,2} \rangle \ .
&\npld
}$$
Clearly, the difference of the superfields on the two sides is BRST closed, and a component evaluation
confirms \npld. In fact, any BRST-closed and local expression obtained from permutation of
$T_{12,3|4,5}$, $k^1_m T^m_{1,2,3|4,5}$ or $k^4_m T^m_{1,2,3|4,5}$ is checked to have vanishing
components, i.e. any element in the BRST cohomology formed from the above two-loop building blocks requires kinematic
poles.

The above numerators $N^{(a)},\ldots,N^{(f)}$ can be mapped to the four-dimensional counterparts of
\CarrascoMN\ by identifying each permutation of $T_{12,3|4,5}$ and $T^m_{1,2,3|4,5}$ with certain
spinor helicity expressions specified in the appendix \appFOUR.
This mapping is only of formal nature and does not result from a (straightforwardly 
applicable) dimensional reduction and specialization to four-dimensional spinor-helicity 
polarizations: The ten-dimensional superspace numerators are local expressions of 
polarizations and momenta
while the spinor-helicity expressions of \CarrascoMN\ contain highly non-local
inverse Gram determinant factors.

\subsec Non-planar two-loop amplitudes in SYM

\noindent
The color-dressed SYM amplitude follows from assembling the six topologies of cubic diagrams
depicted in \figoverview\ \CarrascoMN. In a shorthand notation for the propagators
\eqnn\allints
$$\eqalignno{
I^{(a)}_{1,2,3,4,5}&\equiv  { 1 \over \ell^2 r^2 (\ell+r)^2 (\ell-k_1)^2 (\ell-k_{12})^2 (\ell-k_{123})^2 \, (r-k_5)^2 (r-k_{45})^2}
\cr
I^{(b)}_{1,2,3,4,5}&\equiv  { 1 \over \ell^2 r^2  (\ell+r)^2 (\ell-k_1)^2 (\ell-k_{12})^2 (\ell-k_{123})^2 \, (r-k_5)^2 (\ell+r+k_{4})^2}
\cr
I^{(c)}_{1,2,3,4,5}&\equiv  { 1 \over \ell^2 r^2  (\ell+r)^2(\ell-k_1)^2 (\ell-k_{12})^2  \, (r-k_3)^2(r-k_{34})^2 (\ell+r+k_{5})^2}
&\allints
\cr
I^{(d)}_{1,2,3,4,5}&\equiv  { 1 \over k_{12}^2 \ell^2 r^2  (\ell+r)^2 (\ell-k_{12})^2 (\ell-k_{123})^2 \, (r-k_5)^2 (r-k_{45})^2 }
\cr
I^{(e)}_{1,2,3,4,5}&\equiv  { 1 \over k_{12}^2 \ell^2 r^2  (\ell+r)^2 (\ell-k_{12})^2 (\ell-k_{123})^2 \, (r-k_5)^2 (r+\ell+k_{4})^2}
\cr
I^{(f)}_{1,2,3,4,5}&\equiv  { 1 \over k_{12}^2  \ell^2 r^2  (\ell+r)^2(\ell-k_{12})^2 (\ell+r+k_{3})^2 \, (r-k_5)^2 (r-k_{45})^2} \ ,
}$$
we have
\eqnn\delttt
$$\eqalignno{
{\cal A}^{2-{\rm loop}}_5(\ell,r)&=
{1\over 2}  N^{(a)}_{1,2,3|4,5}(\ell)\, C^{(a)}_{12345}  \, I^{(a)}_{1,2,3,4,5}
+ {1\over 4} N^{(b)}_{1,2,3|4,5}(\ell)\, C^{(b)}_{12345} \, I^{(b)}_{1,2,3,4,5} \cr
&+ {1\over 4} N_{1,2|4,3|5}^{(c)}(\ell,r)\, C^{(c)}_{12345} \, I^{(c)}_{1,2,3,4,5}
+ {1\over 2} N^{(d)}_{12,3|4,5}\, C^{(d)}_{12345} \, I^{(d)}_{1,2,3,4,5} 
 &\delttt \cr
&+ {1\over 4} N^{(e)}_{12,3|4,5}\, C^{(e)}_{12345} \, I^{(e)}_{1,2,3,4,5}
+ {1\over 4} N^{(f)}_{12,3|4,5}\, C^{(f)}_{12345} \, I^{(f)}_{1,2,3,4,5}
+ {\rm sym}(1,2,3,4,5) 
  \ .
}$$
The notation ${\rm sym}(1,2,3,4,5)$ instructs to sum over all the 120 permutations of $(1,2,3,4,5)$,
and the symmetry factors ${1\over 2}$ and ${1\over 4}$ avoid overcounting in the permutation sum
\CarrascoMN. The coefficients of all inequivalent trace configurations in \delttt\ following from
\allcolor\
are checked
to be independently BRST invariant, i.e. the complete color-dressed amplitude \delttt\ is BRST closed.
More details can be found in the appendix \appBRST.

\subsec Two-loop amplitudes in supergravity
\par\subseclab\supgrav

\noindent
Since the color-dressed two-loop five-point SYM amplitude \delttt\ is written in terms of numerators
which satisfy the BCJ color-kinematics duality, the corresponding supergravity integrand is readily
obtained by squaring its numerators \yutin
\eqnn\sugratwo
$$\eqalignno{
{\cal M}^{2-{\rm loop}}_5(\ell,r)&=
{1\over 2}  |N^{(a)}_{1,2,3|4,5}(\ell)|^2  \, I^{(a)}_{1,2,3,4,5}
+ {1\over 4} |N^{(b)}_{1,2,3|4,5}(\ell)|^2 \, I^{(b)}_{1,2,3,4,5} \cr
&+ {1\over 4} |N_{1,2|4,3|5}^{(c)}(\ell,r)|^2 \, I^{(c)}_{1,2,3,4,5}
+ {1\over 2} |N^{(d)}_{12,3|4,5}|^2 \, I^{(d)}_{1,2,3,4,5} 
 &\sugratwo \cr
&+ {1\over 4} |N^{(e)}_{12,3|4,5}|^2 \, I^{(e)}_{1,2,3,4,5}
+ {1\over 4} |N^{(f)}_{12,3|4,5}|^2 \, I^{(f)}_{1,2,3,4,5}
+ {\rm sym}(1,2,3,4,5) 
  \ ,
}$$
where $|N^{(i)}|^2$ is a shorthand for $N^{(i)}\tilde N^{(i)}$ with $i=a,b,c,d,e,f$. The amplitude
\sugratwo\ describes type IIB or type IIA supergravity if the $SO(10)$ Weyl spinors within the
``right-moving'' numerators $\tilde N^{(i)}$ have the same or opposite chirality as compared to the
``left-movers'' $N^{(i)}$. BRST invariance with respect to the left-movers $N^{(i)}$ is inherited from the SYM
amplitude \delttt\ since the accompanying right-moving numerators $\tilde N^{(i)}$ satisfy all the
Jacobi identities of the color factors $C^{(i)}$. The argument extends to the right-moving BRST
variation upon exchange of $N^{(i)}$ and $\tilde N^{(i)}$.

\newsec{UV divergences}
\par\seclab\secfour

\noindent
In this section, we compute the UV divergences of the above two-loop five-point amplitudes in
SYM and supergravity and rewrite the kinematic factors in terms of SYM tree amplitudes. They are
confirmed to match the low-energy limit of the corresponding superstring amplitudes. For completeness
and comparison across different loop-orders, we provide a dimension-agnostic representation of the
one-loop five-point UV divergences in the appendix \appUV.

\subsec Two-loop UV divergences in SYM

In the above BCJ representation of the two-loop five-point SYM amplitude, the kinematic numerators are
at most linear in the loop momentum. Together with the no-triangle property, this implies that the
double-box diagrams dominate in the UV regime of large $\ell^2$. In an expansion around the two-loop
critical dimension $D=7-2\epsilon$ \refs{\MarcusEI}, the planar and non-planar box graphs contribute as
follows in the UV \refs{\BernTwoLoop, \BernTQ},
\eqnn\UVpl
\eqnn\UVnpl
$$\eqalignno{
\int { d^{7-2\epsilon} \ell  \ d^{7-2\epsilon} r  \ (2\pi)^{4\epsilon-14}\over \ell^2 r^2 (\ell+r)^2 \, (\ell-k_1)^2 (\ell-k_{12})^2 \, (r-k_4)^2 (r-k_{34})^2} &= -{\pi + {\cal O}(\epsilon)\over 20(4\pi)^7 \epsilon} \equiv V^{{\rm(P)}} &\UVpl
\cr
\int { d^{7-2\epsilon} \ell  \ d^{7-2\epsilon} r   \ (2\pi)^{4\epsilon-14}\over
\ell^2 r^2 (\ell+r)^2 (\ell+r+k_3)^2 \, (\ell-k_1)^2 (\ell-k_{12})^2 \, (r-k_4)^2}
&= -{\pi + {\cal O}(\epsilon) \over 30(4\pi)^7 \epsilon} \equiv V^{{\rm(NP)}}, &\UVnpl
\cr
}$$
whereas penta-box diagrams are regular in the dimensional regularization parameter $\epsilon$.

The superspace representations for the single-trace subamplitudes
\deli\ and \dels\ yield the following UV divergence in the critical dimension $D=7$:
\eqnn\UVG
$$\eqalignno{
A^{2-{\rm loop}}(1,2,3,4) \, \Big|_{\rm UV} &= V^{{\rm(P)}} \langle T_{1,2|3,4}+T_{4,1|2,3} \rangle=- V^{{\rm(P)}}\langle T_{1,3|2,4} \rangle  &\UVG \cr
A^{2-{\rm loop}}(1,2,3,4,5) \, \Big|_{\rm UV} &= -V^{{\rm(P)}} \Big \langle { T_{12,4|3,5} \over s_{12}} + { T_{23,5|1,4} \over s_{23}} + { T_{1,34|2,5} \over s_{34}} +{  T_{2,45|1,3} \over s_{45}}+{  T_{51,3|2,4} \over s_{51}} 
  \Big \rangle 
}$$
The associated counterterm is the supersymmetrized operator ${\rm Tr}(D^2F^4+F^5)$ which also finds
appearance in the tree-level effective action of the open superstring at order $\ap^3$ \refs{\BilalHB ,
\MedinaNK}. Superspace arguments of \mafraids\ and a component evaluation via \PSS\ confirm that the
kinematic factors in \UVG\ are related to SYM tree amplitudes via
\eqnn\UVF
$$\eqalignno{
A^{2-{\rm loop}}(1,2,3,4) \, \Big|_{\rm UV} &= V^{{\rm(P)}}  s_{12} s_{13} s_{23} A^{{\rm tree}}(1,2,3,4)  &\UVF \cr
 \pmatrix{A^{2-{\rm loop}}(1,2,3,4,5) \cr
 A^{2-{\rm loop}}(1,3,2,4,5) }
\, \Big|_{\rm UV} &= V^{{\rm(P)}}  M_3 \cdot \pmatrix{ A^{\rm tree}(1,2,3,4,5)\cr A^{\rm tree}(1,3,2,4,5)}   \ .
}$$
The $2\times 2$ matrix $M_3$ has been introduced in \motivic\ to describe the momentum dependence of
the $\alpha'$-corrections in open superstring tree-amplitudes \treebbI\ (see also
\refs{\medina,\stieFive}). Its entries are given by
\eqnn\ininKLT
$$\eqalign{
 M_3 &\equiv \pmatrix{m_{11} & m_{12} \cr
                      m_{21} & m_{22}}
}\,,\quad\eqalign{
m_{12} &= - s_{13} s_{24} (s_1+s_2+s_3+s_4+s_5)\hskip88pt\ininKLT \cr
 m_{11} &= s_3 [ - s_1 (s_1+2s_2+s_3)+s_3s_4+s_4^2 ]+s_1s_5 (s_1+s_5)
}$$
with $m_{21} = m_{12} \big|_{2\leftrightarrow 3}$ and $m_{22} = m_{11} \big|_{2\leftrightarrow 3}$ as
well as $s_i \equiv s_{i,i+1}$ subject to $s_{5}= s_{15}$. As a main virtue of writing the UV
divergence \UVF\ in terms of $A^{{\rm tree}}(\ldots)$, it is agnostic to the choice of kinematic
variables and can be adapted to any spacetime dimension by choosing the appropriate representation of
the SYM tree.

Upon combination with the non-planar sector, the color-dressed amplitude \delttt\ for traceless
gauge group generators ${\rm Tr}(t^i)=0$ yields the UV divergence
\eqnn\UVnonplan
$$\eqalignno{
&{\cal A}^{2-{\rm loop}}_5 \, \Big|_{{\rm UV}} =  \big[ V^{{\rm(P)}} N_c^2 +  12 (V^{{\rm(P)}}+V^{{\rm(NP)}}) \big]\cr
& \ \ \ \ \ \ \times \Big\{
\pmatrix{{\rm Tr}(t^1t^2t^3t^4t^5)\cr
         {\rm Tr}(t^1t^3t^2t^4t^5)}^{\! T}\!\!\! \cdot  M_3 \cdot
\pmatrix{ A^{\rm tree}(1,2,3,4,5)\cr
          A^{\rm tree}(1,3,2,4,5)} + (2,3|2,3,4,5) \Big\} &\UVnonplan \cr
& \ \ - 12 N_c (V^{{\rm(P)}}+V^{{\rm(NP)}}) \Big\{ {\rm Tr}(t^1 t^2 t^3) {\rm Tr}(t^4 t^5) \cr
& \ \ \ \ \ \ \times  s_{45}^2 \big[ s_{24} A^{\rm tree}(1,3,2,4,5) -s_{34} A^{\rm tree}(1,2,3,4,5) \big] + (4,5|1,2,3,4,5) \Big\} \cr
}$$
after expanding the color factors $C_i$ in a trace basis. The notation $(a_1,a_2|a_1,a_2,\ldots, a_n) $ instructs to sum over all possible ways to choose two
elements out of the set $(a_1,a_2,\ldots,a_n)$, for a total of ${n \choose 2}$ terms. The 
coefficients of the multitrace color structure ${\rm Tr}(t^1 t^2 t^3) {\rm Tr}(t^4 t^5)$ stem from 
the kinematic factor
\eqn\UVnonpltwo{
\Big \langle {T_{12,3|4,5} \over s_{12} }+{T_{23,1|4,5} \over s_{23}} +{T_{31,2|4,5} \over s_{13} }
\Big \rangle=
s_{45}^2 \big[ s_{24} A^{\rm tree}(1,3,2,4,5) -s_{34} A^{\rm tree}(1,2,3,4,5) \big]  \ ,
}
which reduces to the four-dimensional expression $s_{45} \big( {\gamma_{12} \over s_{12}} +
{\gamma_{23} \over s_{23}} +{\gamma_{31} \over s_{31}} \big)$ in \CarrascoMN\ under the mapping \jja.
Similarly, the single-trace kinematic factor in \UVG\ is mapped to the spinor helicity expression
$\beta_{12345}+ {\gamma_{12} \over s_{12}}(s_{35}-2s_{12}) + {\rm cyc}(1,2,3,4,5)$ from \CarrascoMN\
under \jja.

\subsec Two-loop UV divergences in supergravity

According to their BCJ construction, the penta-box numerators in the supergravity amplitude \sugratwo\
involve up to two powers of loop momentum whereas double-box numerators remain independent on $\ell$.
Accordingly, the leading UV contributions in $D=7-2\epsilon$ dimensions are given by
\eqnn\UVH
\eqnn\UVfive
$$\eqalignno{
{\cal M}^{2-{\rm loop}}_4 \, \Big|_{\rm UV} &= 2 (V^{{\rm(P)}}+V^{{\rm(NP)}}) \Big\{ \big| \langle T_{1,2|3,4} \rangle \big|^2 +\big| \langle T_{1,3|2,4} \rangle \big|^2+\big| \langle T_{1,4|2,3} \rangle \big|^2 \Big\}
&\UVH\cr
{\cal M}^{2-{\rm loop}}_5 \, \Big|_{\rm UV} &= 2 (V^{{\rm(P)}}+V^{{\rm(NP)}})  \Big\{
{\big|\langle T_{12,3|4,5}\rangle \big|^2\over s_{12}} + {\big|\langle T_{12,4|3,5}\rangle \big|^2\over s_{12}}
+ {\big|\langle T_{12,5|3,4}\rangle \big|^2\over s_{12}} \cr
& \ \ \ \ \ \ \ \ \ \ \ \ \ \ \ \ \ \ \ \ \ \ \ \ \ \ \ \ \ \ \ \ \ +
\big|\langle T^m_{3,4,5|1,2}\rangle\big|^2 +
(1,2|1,2,3,4,5) \Big\} \,. &\UVfive 
}$$
As a consistency check of the proposed supergravity amplitude \sugratwo, the above UV divergences have
to agree with the low-energy limits of the two-loop closed superstring amplitudes \GreenYU. Indeed, the
kinematic factors in \UVH\ emerge in the low-energy limits at four-points \twoloop\ and
five-points\foot{The five-point superstring computation in \twoloopfive\ leads to a different
representation of $T_{A,B|C,D}$ and $T^m_{1,2,3|4,5}$ in terms of the non-minimal pure spinor variables
\NMPS. However, BRST-invariant expressions do not depend on the representation of their composing building block.}
\twoloopfive. The polarization dependence can be written in terms of SYM tree subamplitudes: At four
points, the superspace arguments of \mafraids\ imply that
\eqn\UVI{
\big| \langle T_{1,2|3,4} \rangle \big|^2 + \big| \langle T_{1,3|2,4} \rangle
\big|^2 + \big| \langle T_{1,4|2,3} \rangle \big|^2 =
(s_{12}^2 + s_{13}^2 + s_{23}^2) \big| s_{12} s_{23} A^{\rm tree}(1,2,3,4) \big|^2\,,
}
and a {\it ten-dimensional\/} type IIB component evaluation at five points yields \twoloopfive
\eqnn\UVJ
$$\eqalignno{
&{\big|\langle T_{12,3|4,5}\rangle \big|^2\over s_{12}} + {\big|\langle T_{12,4|3,5}\rangle \big|^2\over s_{12}}
+ {\big|\langle T_{12,5|3,4}\rangle \big|^2\over s_{12}} + \big|\langle T^m_{3,4,5|1,2}\rangle\big|^2 +
(1,2|1,2,3,4,5) \, \Big|^{D=10}_{{\rm IIB}} \cr
&=2\pmatrix{\tilde A^{\rm tree}(1,2,3,5,4)\cr
 \tilde A^{\rm tree}(1,3,2,5,4)}^{T} \cdot S_0 \cdot M_5 \cdot \pmatrix{ A^{\rm tree}(1,2,3,4,5)\cr A^{\rm tree}(1,3,2,4,5)}  
  \times 
  \cases{\ \ \, 1 \,  \ : \ \ h^5 \cr -{3\over 5} \ : \ \phi h^4} 
  \ . &\UVJ
}$$
The $2\times2$ matrix $M_5$ with entries of order $s_{ij}^5$ has been introduced in \motivic\ to
describe the $(\ap)^5$-correction to the open five-point superstring tree-level amplitude \treebbI\ and
can be downloaded from the website \WWWalpha. The matrix
\eqn\momker{
 S_0 \equiv  \pmatrix{s_{12}(s_{13}+s_{23}) & s_{12}s_{13}\cr
 s_{12}s_{13} & s_{13}(s_{12}+s_{23})} 
 }
captures the Mandelstam invariants in the field-theory limit of the KLT relations \KLTref. The
shorthands $h^5$ and $\phi h^4$ in \UVJ\ refer to type IIB components with zero and two units of
R-symmetry charge such as five gravitons $h^5$ or four gravitons and one (axio-)dilaton $\phi
h^4$, respectively.

\subsec UV divergence and R-symmetry
\par\subseclab\UVRsym

\noindent As seen in \UVfive\ and \UVJ, the UV divergence of the supergravity two-loop five-point amplitude
is given by the same superspace expression that arises in the
low-energy limit of the corresponding closed-string amplitude computed in \twoloopfive.
Furthermore, the string amplitude for R-symmetry violating states such as
$\phi h^4$ does not vanish; its characteristic coefficient $-3/5$ in \UVJ\ agrees with expectations from
S-duality considerations for the type IIB string \refs{\Rviolating,\twoloopfive}. These facts give rise to worry that the
two-loop UV divergence in supergravity might violate the R-symmetry as well.

However, that is not the case\foot{We thank John-Joseph Carrasco and Henrik Johansson for helpful email
correspondence on this point.}. To see this, note that the two-loop UV divergence of supergravity
occurs in the critical dimension $D=7$ whereas the string $\phi h^4$ amplitude \UVJ\ is computed in
$D=10$. Furthermore, recall that the graviton polarization $h_{mn}$ is the traceless part of $e_{(m}
\tilde e_{n)}$ while the dilaton wavefunction $(\d_{mn} - k_m \bar k_n - k_n\bar k_m)\phi$ covers the
trace part with respect to the little group whose reference momentum $\bar k_m$ satisfies $\bar k\cdot \bar k=0$
and $k\cdot \bar k = 1$ \GrossSloan. Care must be taken when amplitudes involving dilatons are computed
in general dimensions $D$, since the dimensional reduction of the little group trace yields
\eqn\dilatonD{
e\cdot \tilde e = (D-2)\phi\, \ .
}
Note that the four-dimensional dilaton state is tied to R-symmetry anomalies in $D=4$ supergravities with ${\cal N}\leq 4$ supersymmetry, see e.g. \CarrascoYPA.

In fact, using the component form of the building blocks $T_{12,3|4,5}$ and 
$T^m_{1,2,3|4,5}$ available to download in \WWW\ one can check that the kinematic factor \UVJ\ in $D$ dimensions becomes\foot{The dimensional reduction of this component calculation
is performed {\it after} expanding the contracted ten-dimensional Levi-Civita bilinears
$\varepsilon^{m n_1n_2\ldots n_9} \varepsilon_{m p_1 p_2\ldots p_9} =
- 9! \delta^{[n_1}_{p_1} \delta^{n_2}_{p_2} \cdots \delta^{n_9]}_{p_9}$
due to the contractions $ \langle T^m_{ \ldots} \rangle \langle \tilde T^m_{ \ldots} \rangle$ between
left- and right-moving superfields. If state 1 is chosen to be a $D$-dimensional dilaton, then the
only dependence on $D$ stems from $e_1\cdot \tilde e_1 = (D-2)\phi$.}
\eqnn\dilatoninD{
$$\eqalignno{
&{\big|\langle T_{12,3|4,5}\rangle \big|^2\over s_{12}} + {\big|\langle T_{12,4|3,5}\rangle \big|^2\over s_{12}}
+ {\big|\langle T_{12,5|3,4}\rangle \big|^2\over s_{12}} + \big|\langle T^m_{3,4,5|1,2}\rangle\big|^2 +
(1,2|1,2,3,4,5) \, \Big|^{D}_{{\rm IIB}} \cr
&=2\pmatrix{\tilde A^{\rm tree}(1,2,3,5,4)\cr
 \tilde A^{\rm tree}(1,3,2,5,4)}^{T} \cdot S_0 \cdot M_5 \cdot \pmatrix{ A^{\rm tree}(1,2,3,4,5)\cr A^{\rm tree}(1,3,2,4,5)}  
  \times 
  \cases{\ \ \ \ \  1 \,  \ \ \ \ \ : \ \ h^5 \cr  \cr \displaystyle {(7-D)\over 5} \ : \ \phi h^4} 
  \ . &\dilatoninD
}$$
Therefore, the $\phi h^4$ contribution vanishes in the critical dimension $D=7$ relevant for 
the two-loop supergravity UV divergence, and the R-symmetry violation is circumvented.


\newsec{Conclusion and outlook}
\par\seclab\secfive

\noindent In this paper the two-loop five-point amplitudes of both SYM and type II supergravity in ten
dimensions were computed using the BRST cohomology method of \refs{\towards,\MafraJQ, \MafraGJA}. 
The supersymmetric kinematic numerators are manifestly local, and their derivation follows an intuitive
mapping between cubic graphs and superspace building blocks as guided by their BRST
variation. Inspired by the BCJ-satisfying four-dimensional representation of \CarrascoMN, 
ten-dimensional numerators for all the planar and non-planar diagrams were written down in a form compatible with the
color-kinematics duality.

The compatibility of the BRST principle \principle\ with the color-kinematics duality has already been
encountered for tree-level $n$-point numerators \MafraKJ\ and one-loop five-point numerators \MafraGJA.
Both of these cases emerge naturally from the field-theory limit of the corresponding superstring
amplitudes, in the same way as the resulting BCJ subamplitude relations at tree-level \BCJ\ have an
elegant derivation from string theory \refs{\BjerrumBohrRD,\StiebergerHQ}. This suggests that the
superstring is a convenient starting point to understand the duality between color and kinematics in a
broader context, see e.g. \OchirovXBA\ for an example at reduced supersymmetry.

The string theory derivation of this work's results is an open problem since the genus-two five-point
worldsheet correlator in \twoloopfive\ was determined only in the low-energy limit. Still, the
kinematic building blocks $T_{12,3|4,5}$ and $T^m_{1,2,3|4,5}$ have an alternative representation in
\twoloopfive\ in variables of the non-minimal pure spinor formalism \NMPS\ which gives rise to the same
component expansions when combined in a BRST-invariant manner. In particular, their appearance in the
UV divergence \UVH\ of the supergravity amplitude and the low-energy limit of the closed superstring is
identical, confirming the general expectation of \GreenYU. Once the completion of the correlator in
\twoloopfive\ beyond the low-energy limit is achieved, it would be desirable to reproduce the present
field-theory amplitudes, using for example the techniques of \tropical. Also, a derivation from the
non-minimal pure spinor version of the ambitwistor string \AdamoHOA\ would be desirable.

It would also be interesting to study the higher-point construction of the two-loop
SYM amplitudes. In this case, a sequence of BRST-covariant tensorial building blocks is required to
describe higher powers of loop momentum in $(n\geq 6)$-gon subdiagrams. At one loop, the
analogous tensors have been found in \MafraGSA\ and used in the BRST cohomology derivation of the
six-point one-loop SYM amplitude in \MafraGJA. Furthermore, the general form of the BRST principle
\principle\ motivates to assemble higher-loop amplitudes in the same manner as described in this paper.
The four-point BCJ representations at three and four loops in \BCJloop\ and \BernUF\ are expected to
provide valuable guidance. For the design of superspace numerators, the superfields of higher-mass
dimensions constructed in \HighSYM\ will play an essential role, and the low-energy limit of the
three-loop superstring amplitude in \threeloop\ constrains the leading $\ell$-dependence in the
numerators.

\bigskip \noindent{\bf Acknowledgements:} We thank Piotr Tourkine for valuable discussions 
and John-Joseph Carrasco as well as Henrik Johansson for helpful email correspondence. 
We are grateful to Piotr Tourkine for useful comments on the draft and to Humberto Gomez 
for collaboration on related topics. We acknowledge support by the European
Research Council Advanced Grant No. 247252 of Michael Green. We would like to thank the 
organizers of the workshop ``Superstring Perturbation Theory'' held at Perimeter Institute in 
April 2015 for the opportunity to present this work. This research was supported in part by
Perimeter Institute for Theoretical Physics. Research at Perimeter Institute is supported by 
the Government of Canada through Industry Canada and by the Province of Ontario through the ministry of
Economic Development \& Innovation. OS is grateful to DAMTP for kind hospitality during initial steps
of this work.

\appendix{A}{Diagrammatic bookkeeping of BRST variations}
\applab\appBRST

\noindent In order to verify the BRST invariance of the integrands \dels\ and \delttt, it is convenient
to devise a diagrammatic bookkeeping for the effect of the numerators' BRST variation on their
associated loop integrals. This exploits the central requirement \principle\ that each summand in the
BRST variation of a particular numerator must contain an inverse propagator of its associated integral.
The need of redefining loop momenta to see the cancellation among different terms is bypassed once the
remaining propagators are represented by a graph and manipulated through its automorphism symmetries.

To understand how this comes about, recall the BRST variation of the planar penta-box
numerators $N^{(a)}_{1,2,3|4,5}(\ell)$ defined in \Nadef,
\eqnn\recallQNa
$$\eqalignno{
QN^{(a)}_{1,2,3|4,5}(\ell) &=  {1\over 2} \Big( V_1 T_{2,3|4,5} \big[ \ell^2 - (\ell-k_{1})^2 \big]
+V_2 T_{1,3|4,5} \big[ (\ell-k_{1})^2 - (\ell-k_{12})^2 \big] \cr
&\ \ \ \ \ \ \ \ \   +V_3 T_{1,2|4,5} \big[ (\ell-k_{12})^2 - (\ell-k_{123})^2 \big] \Big) \,.
&\recallQNa
}$$
Therefore, the combined effect of the BRST variation on the product of $N^{(a)}$ and its 
associated penta-box propagators is given by
\eqnn\monster
$$\eqalignno{
&{2QN^{(a)}_{1,2,3|4,5}(\ell)\over \ell^2 r^2 (\ell+r)^2  (\ell-k_1)^2 (\ell-k_{12})^2 (\ell-k_{123})^2\, (r-k_5)^2 (r-k_{45})^2}  \cr
&=  {1\over r^2 (\ell+r)^2  (r-k_5)^2 (r-k_{45})^2} \times \Big\{ {V_1 T_{2,3|4,5}\over   (\ell-k_1)^2 (\ell-k_{12})^2 (\ell-k_{123})^2},&\monster \cr
& \ \ \ \ \ \ + {- V_1 T_{2,3|4,5} + V_2T_{1,3|4,5}\over \ell^2   (\ell-k_{12})^2 (\ell-k_{123})^2 } + {- V_2 T_{1,3|4,5} + V_3 T_{1,2|4,5}\over \ell^2   (\ell-k_{1})^2 (\ell-k_{123})^2 } + {- V_3 T_{1,2|4,5} \over \ell^2   (\ell-k_{1})^2 (\ell-k_{12})^2 } \Big\} \ , \cr
}$$
whose diagrammatic interpretation is given in the following figure:
\medskip
\centerline{{\epsfxsize=0.85\hsize\epsfbox{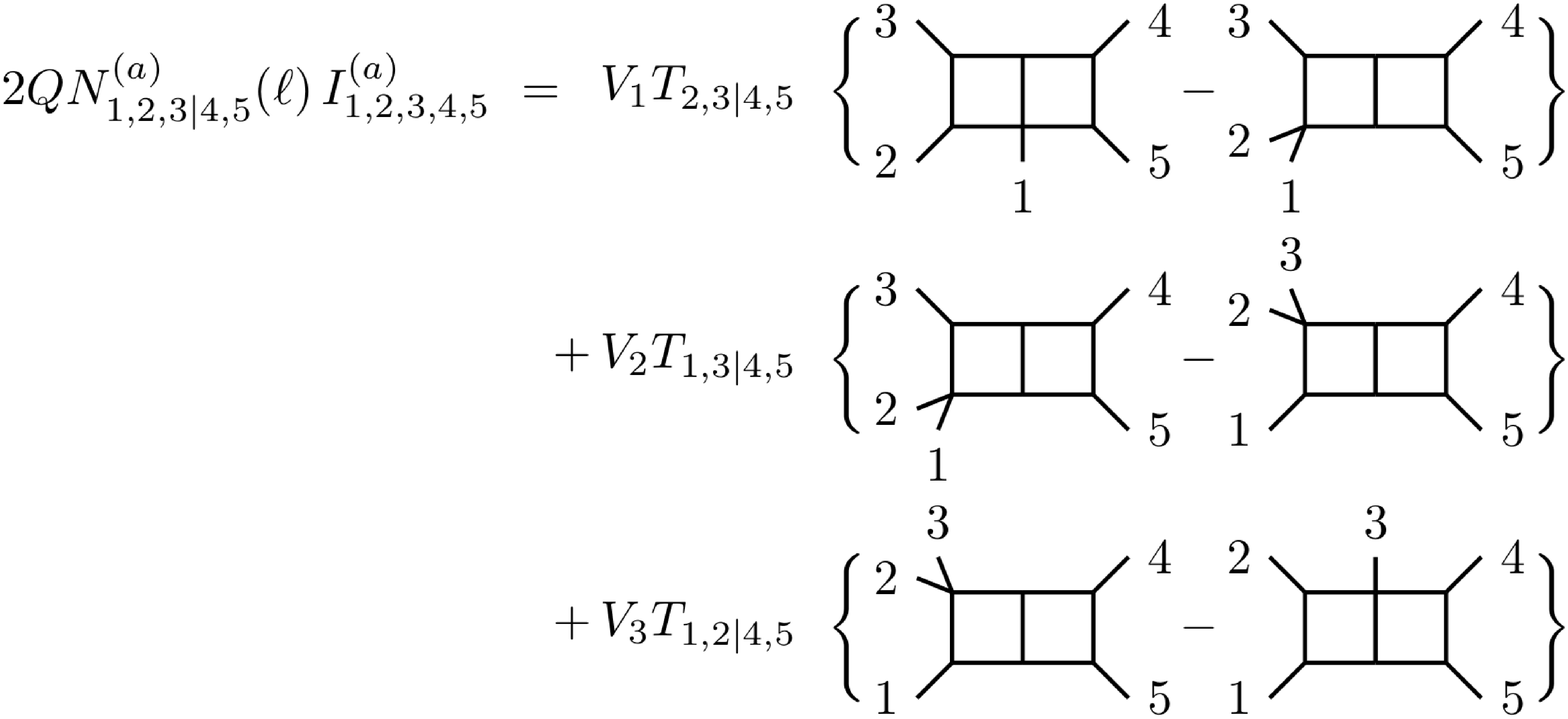}}}
\medskip
\noindent So it is clear that out of the six terms in \monster\ only two distinct integral topologies occur.

\ifig\figtops{Five topologies of integrals occurring in the BRST variation of individual graphs}
{\epsfxsize=0.90\hsize\epsfbox{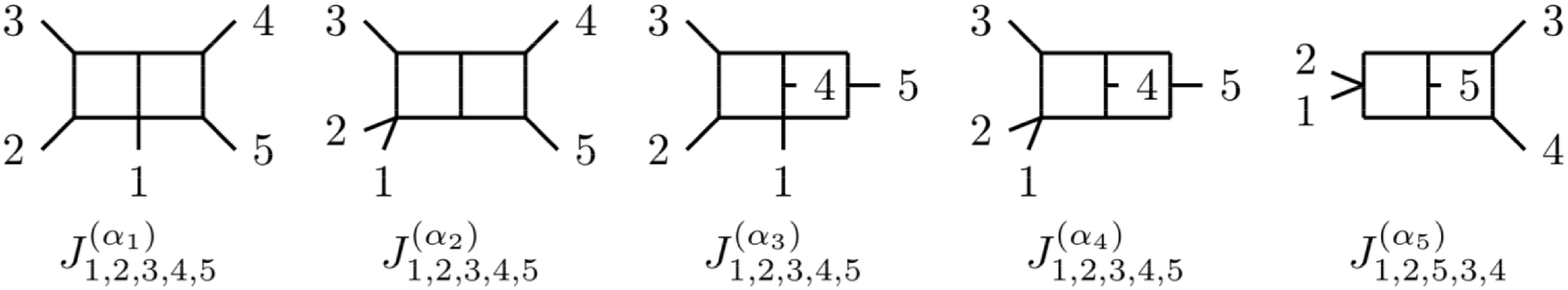}}

\noindent Together with remaining graphs in \figoverview, the BRST variation of the color-dressed
five-point two-loop amplitude in \delttt\ turns out to be captured by the five integral topologies
$J^{(\alpha_1)},\ldots,J^{(\alpha_5)}$ depicted in \figtops. Given the shorthands
$I^{(a)},\ldots,I^{(f)}$ in \allints, they arise from the following BRST variations
\eqnn\BRSTall
$$\eqalignno{
2 Q N^{(a)}_{1,2,3|4,5}(\ell) I^{(a)}_{1,2,3,4,5}&=
         V_1 T_{2,3|4,5}\big(
        J^{(\a_1)}_{1,2,3,4,5}
        - J^{(\a_2)}_{1,2,3,4,5}
        \big)
        \cr
     + V_2 &T_{1,3|4,5}\big(
        J^{(\a_2)}_{1,2,3,4,5}
        - J^{(\a_2)}_{2,3,1,5,4}
        \big)   + V_3 T_{1,2|4,5}\big(
        J^{(\a_2)}_{2,3,1,5,4}
        - J^{(\a_1)}_{3,4,5,1,2}
        \big)
	\cr
2 Q N^{(b)}_{1,2,3|4,5}(\ell) I^{(b)}_{1,2,3,4,5}&=
         V_1 T_{2,3|4,5}\big( J^{(\a_3)}_{1,2,3,4,5}
        -  J^{(\a_4)}_{1,2,3,4,5}
	\big)
        \cr
  + V_2 &T_{1,3|4,5}\big( J^{(\a_4)}_{1,2,3,4,5}
        - J^{(\a_4)}_{2,3,1,4,5}
	\big)      + V_3 T_{1,2|4,5}\big( J^{(\a_4)}_{2,3,1,4,5}
        - J^{(\a_3)}_{3,2,1,4,5}
        \big)\cr
2 Q N^{(c)}_{1,2|4,3|5}(\ell,r) I^{(c)}_{1,2,3,4,5}&=
         V_1 T_{2,5|3,4}\big( J^{(\a_3)}_{1,3,4,5,2}
        -  J^{(\a_5)}_{1,2,5,4,3}
	\big)
       \cr
  + V_2 &T_{1,5|3,4}\big( J^{(\a_5)}_{1,2,5,4,3}
        -  J^{(\a_3)}_{2,4,3,5,1}
	\big)       + V_3 T_{4,5|1,2}\big( J^{(\a_3)}_{3,1,2,5,4}
        -  J^{(\a_5)}_{3,4,5,2,1}
	\big) \cr
        + V_4 &T_{3,5|1,2}\big( J^{(\a_5)}_{3,4,5,2,1}
        - J^{(\a_3)}_{4,2,1,5,3}
	\big)  + V_5 T_{1,2|3,4}\big( J^{(\a_1)}_{5,4,3,1,2}
        - J^{(\a_1)}_{5,1,2,4,3}
	\big)\cr
2 Q N^{(d)}_{12,3|4,5} I^{(d)}_{1,2,3,4,5}&=
        (V_1 T_{2,3|4,5} - V_2 T_{1,3|4,5})\, J^{(\a_2)}_{1,2,3,4,5} \cr
2 Q N^{(e)}_{12,3|4,5} I^{(e)}_{1,2,3,4,5}&=
        ( V_1 T_{2,3|4,5}    - V_2 T_{1,3|4,5})\, J^{(\a_4)}_{1,2,3,4,5}&\BRSTall\cr
2 Q N^{(f)}_{12,3|4,5} I^{(f)}_{1,2,3,4,5}&=
         (V_1 T_{2,3|4,5} - V_2 T_{1,3|4,5})\, J^{(\a_5)}_{1,2,3,4,5} \ ,
}$$
see \figQNb\ and \figdouble\ for a diagrammatic illustration of these identities.
Using the BRST variations \BRSTall, the kinematic identities \twoloC\ and the
automorphism symmetries
\eqnn\autosym
$$\eqalignno{
J^{(\a_1)}_{1,2,3,4,5} &= J^{(\a_1)}_{1,5,4,3,2}\,,\quad
J^{(\a_2)}_{1,2,3,4,5} = J^{(\a_2)}_{2,1,3,4,5}\,,\quad
J^{(\a_3)}_{1,2,3,4,5} =  J^{(\a_3)}_{1,2,3,5,4}\,,\quad
\cr
J^{(\a_4)}_{1,2,3,4,5} &=  J^{(\a_4)}_{2,1,3,4,5} =  J^{(\a_4)}_{1,2,3,5,4}\,,\quad
J^{(\a_5)}_{1,2,3,4,5} =  J^{(\a_5)}_{2,1,3,4,5} =  J^{(\a_5)}_{1,2,3,5,4} \,,
&\autosym
}$$
one can show that \delttt\ is BRST closed.

\ifig\figQNb{When multiplied by their corresponding integrands of \allints, the BRST variations of the
non-planar penta-box numerators $N^{(b)}_{1,2,3|4,5}$ and $N^{(c)}_{1,2,3|4,5}$ admit a diagrammatic
interpretation.}
{\hskip-49pt\vbox{%
\epsfxsize=1.05\hsize\epsfbox{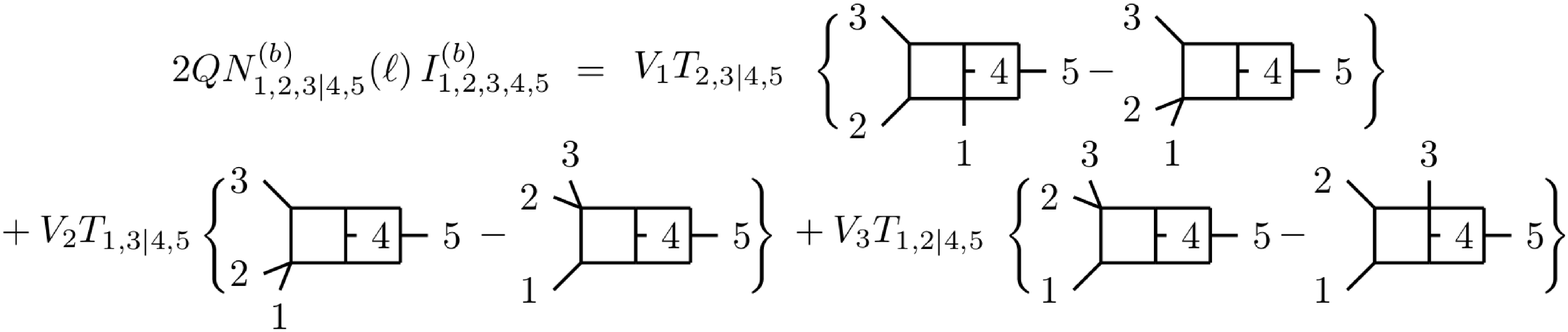}
\smallskip
\epsfxsize=1.05\hsize\epsfbox{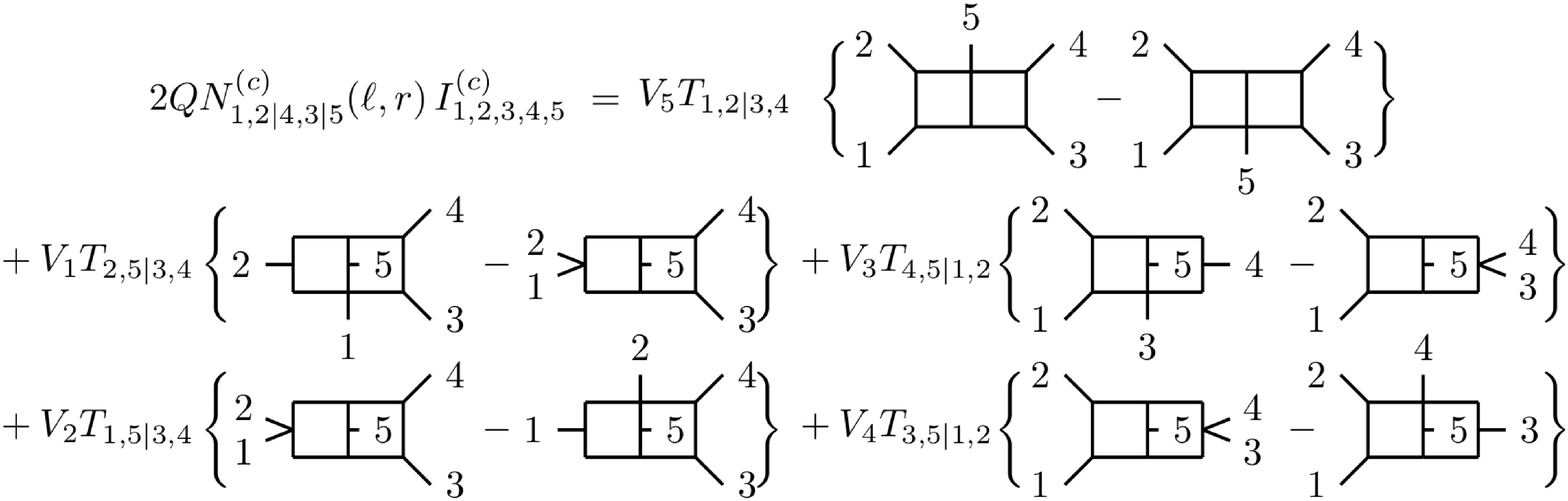}
}}

\ifig\figdouble{The diagrammatic interpretation for the BRST variation of the double-boxes after
multiplication by their corresponding integrands given in \allints.}
{\epsfxsize=0.70\hsize\epsfbox{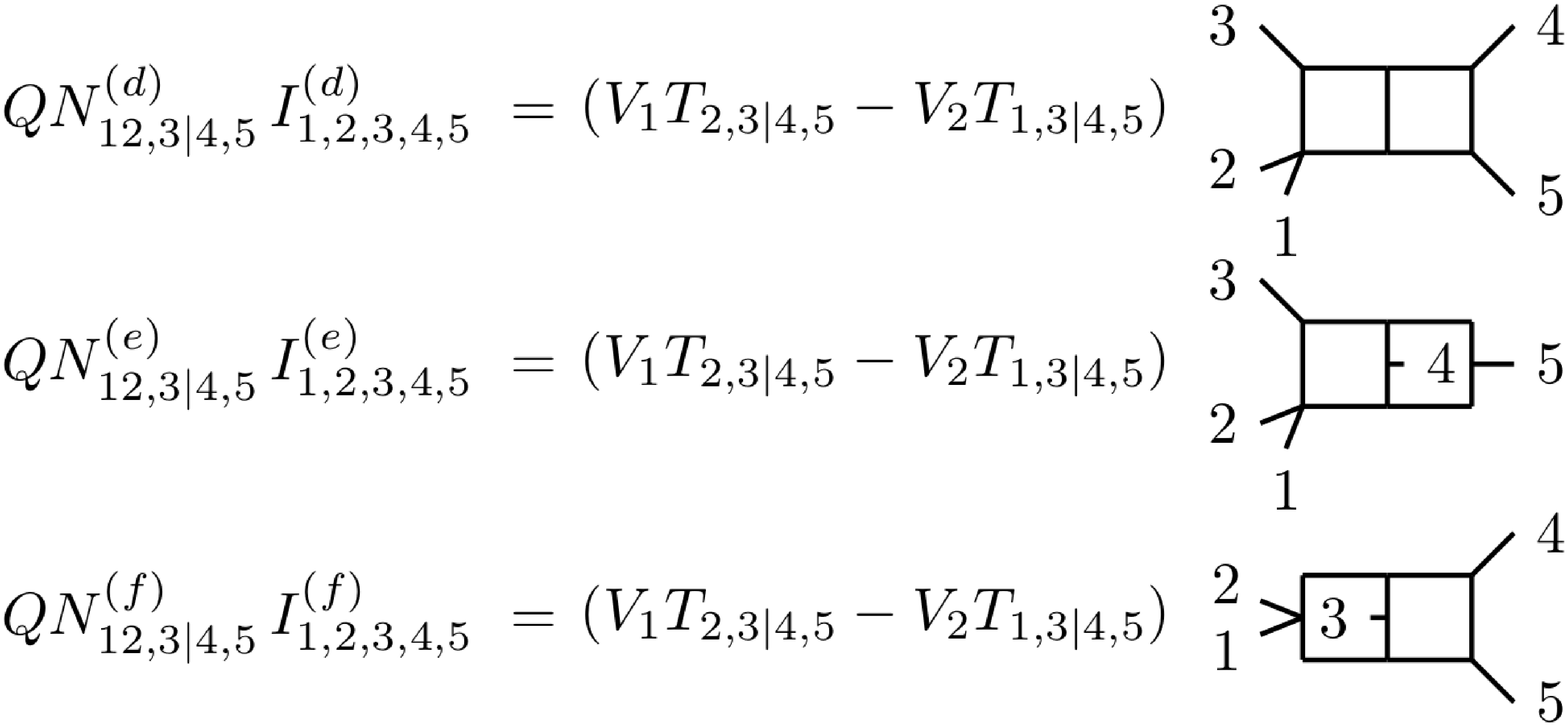}}

\appendix{B}{Comparison with the four-dimensional solution}
\applab\appFOUR

\noindent
Just like the ten-dimensional numerators presented in the main text, the four-dimensional numerators written in Table~1 of \CarrascoMN\ are composed of scalar and vector building blocks. The numerators can be mapped into each other once the building blocks are replaced as\foot{The formal replacement rules in \jja\ do not imply that dimensional reduction of $\langle T_{12,3|4,5} \rangle$ and $\langle T^m_{1,2,3|4,5} \rangle$ yields the combinations of $\gamma_{ij}$ on the right hand sides.}
\eqnn\jja
$$\eqalignno{
T_{12,3|4,5}&\rightarrow \gamma_{12} (s_{45}-{1\over 2}s_{12}) + {1\over 4} s_{12}(\gamma_{23}-\gamma_{13})&\jja \cr
T^m_{1,2,3|4,5}&\rightarrow \gamma_{45}(k_5^m-k_4^m) + {1\over 2}\big[ \gamma_{12} (k_2^m-k_1^m) + \gamma_{23}(k_3^m - k_2^m) + \gamma_{31}(k_1^m-k_3^m) \big] \ , 
}$$
where $\gamma_{ij}$ is built from spinor helicity expressions $[ i \, j ]$ and $\delta^8({\cal Q})$ defined in \CarrascoMN,
\eqn \gammaspin{
\gamma_{12} \equiv \delta^8({\cal Q}) {[1\,2]^2  [3\,4] [3\,5] [4\,5] \over \varepsilon_{\mu \nu \lambda \rho} k_1^\mu k_2^\nu k_3^{\lambda} k_4^{\rho}} \ .
}
The denominator is totally antisymmetric in $1,2,\ldots,5$ and introduces a spurious singularity in the determinant ${\rm Det}( k_i^\mu)$ or the directed volume of the momenta $k_1^\mu, k_2^\nu ,k_3^{\lambda}$ and $k_4^{\rho}$ with four-dimensional vector indices $\mu,\nu,\lambda,\rho$.

Given that this spinor helicity building block $\gamma_{ij}=-\gamma_{ji}$ satisfies
\eqnn\jjc
\eqnn\jjd
$$\eqalignno{
0&=\gamma_{12}+\gamma_{13}+\gamma_{14}+\gamma_{15}&\jjc \cr
0&= (\gamma_{12}+\gamma_{13})(s_{23}-s_{45})+\gamma_{23}(s_{12}-s_{13})+\gamma_{45}(s_{14}-s_{15})
 \ , &\jjd
}$$
one can straightforwardly check that the superspace identities
\eqnn\jjf
$$\eqalignno{
0&= \langle T_{12,3|4,5}+T_{12,4|5,3}+T_{12,5|3,4} \rangle \cr
0&= \langle T_{3,4,5|1,2}^m+T_{2,4,5|1,3}^m+T_{1,4,5|2,3}^m-T_{1,2,3|4,5}^m \rangle  &\jjf \cr
0&= \langle T_{15,2|3,4}+T_{25,1|3,4}+T_{35,4|1,2}+T_{45,3|1,2}-k_m^5(T^m_{3,4,5|1,2}+T^m_{1,2,5|3,4}) \rangle 
}$$
are all respected by the mapping in \jja.

\appendix{C}{One-loop UV divergences}
\applab\appUV

\noindent
At one loop, the four-point UV divergences in SYM and supergravity are well-known from
\refs{\BernTQ, \GreenFT}. Using the pure spinor superspace representation of the five-point one-loop
amplitudes in \MafraGJA\ and the identities in the appendix B of \EOMBBs\ we write the five-point UV divergence
in terms of SYM tree amplitudes.

 \subsec One-loop UV divergences in SYM

The counting of loop momenta is identical in SYM amplitudes at one- and two-loops, hence,
 the box diagrams dominate in the $(8-2\epsilon)$ dimensional UV regime at one-loop via
 \eqn\boxdiv{
 \int {d^{8-2\epsilon}\ell \ (2\pi)^{2\epsilon-8} \over \ell^2(\ell-k_1)^{2}(\ell-k_{12})^{2}(\ell-k_{123})^{2}} = {i+ {\cal O}(\epsilon) \over 6(4\pi)^{4} \epsilon} \ .
 }
The UV divergence in the critical dimension $D=8$ is characterized by a supersymmetrized $F^4$
counterterm \GreenFT. From the pure spinor superspace representations of the four- and five-point
one-loop amplitudes in SYM \MafraGJA, one can extract the UV divergence,
\eqnn\UVA
 $$\eqalignno{
 A^{1-{\rm loop}}(1,2,3,4) \, \Big|_{\rm UV} &= {i\over 6(4\pi)^{4} \epsilon} \langle V_1 T_{2,3,4} \rangle  &\UVA \cr
 A^{1-{\rm loop}}(1,2,3,4,5) \, \Big|_{\rm UV} &= {i\over 6(4\pi)^{4} \epsilon}  \Big \langle { V_{12} T_{3,4,5} \over s_{12}} + { V_{1} T_{23,4,5} \over s_{23}} + { V_{1} T_{2,34,5} \over s_{34}} +{ V_{1} T_{2,3,45} \over s_{45}}+{ V_{51} T_{2,3,4} \over s_{51}}  
   \Big \rangle \ ,
 }$$
see \MafraGJA\ for the scalar one-loop building blocks $T_{A,B,C}$. This reproduces the pure spinor
analysis of $F^4$ amplitudes in \oneloopbb, and the same matching can be found for the six-point
one-loop amplitudes in \MafraGJA. An equivalent representation in terms of SYM tree amplitudes,
\eqnn\UVB
 $$\eqalignno{
 A^{1-{\rm loop}}(1,2,3,4) \, \Big|_{\rm UV} &= {i\over 6(4\pi)^{4} \epsilon}  s_{12} s_{23} A^{{\rm tree}}(1,2,3,4)  &\UVB \cr
  \pmatrix{A^{1-{\rm loop}}(1,2,3,4,5) \cr
 A^{1-{\rm loop}}(1,3,2,4,5) }
\, \Big|_{\rm UV} &= {i\over 6(4\pi)^{4} \epsilon} P_2 \cdot \pmatrix{ A^{\rm tree}(1,2,3,4,5)\cr A^{\rm tree}(1,3,2,4,5)}  
  \ ,
 }$$
reproduces the $(\ap)^2$-correction to superstring tree amplitudes \treebbI\ with $P_2$ defined by \motivic
\eqn\Ptwo{
P_2 \equiv 
\pmatrix{ s_{12} s_{34}-s_{34}s_{45}-s_{51}s_{12} & s_{13}s_{24} \cr
s_{12}s_{34} & s_{13} s_{24}-s_{24}s_{45}-s_{51}s_{13}}  \ .
}
Together with the non-planar sector, the color-dressed one-loop amplitude with traceless gauge group
generators ${\rm Tr}(t^i)=0$ gives rise to the UV divergence
\eqnn\UVnonpl
$$\eqalignno{
&{\cal A}^{1-{\rm loop}}_5 \, \Big|_{{\rm UV}} =  {iN_c\over 6(4\pi)^{4} \epsilon}
\pmatrix{{\rm Tr}(t^1t^2t^3t^4t^5)\cr {\rm Tr}(t^1t^3t^2t^4t^5)}^T \cdot  P_2 \cdot \pmatrix{ A^{\rm tree}(1,2,3,4,5)\cr A^{\rm tree}(1,3,2,4,5)} + (2,3|2,3,4,5)  &\UVnonpl \cr
& + {i \over (4\pi)^{4} \epsilon}   {\rm Tr}(t^1 t^2 t^3) {\rm Tr}(t^4 t^5) s_{45} \big[ s_{24} A^{\rm tree}(1,3,2,4,5) -s_{34} A^{\rm tree}(1,2,3,4,5) \big]  + (4,5|1,2,3,4,5)  \ .
}$$
The kinematic factor along with the multitrace
\eqn\Coneloop{
\Big \langle {V_1 T_{23,4,5} \over s_{23}} + {V_{12} T_{3,4,5} \over s_{12}} + {V_{31} T_{2,4,5} \over s_{13}} \Big \rangle =
s_{45} \big[ s_{24} A^{\rm tree}(1,3,2,4,5) -s_{34} A^{\rm tree}(1,2,3,4,5) \big] 
}
closely resembles the superfield structure of \UVnonpltwo\ and was denoted by $C_{1|23,4,5}$ in \refs{\oneloopbb,\EOMBBs}.

 \subsec One-loop UV divergences in supergravity

The UV behavior of $n$-point supergravity amplitudes at one loop is affected by any $p$-gon diagram
with $4\leq p\leq n$. At four- and five-points, the leading UV divergence in dimensions $D= 8-2\epsilon$
can be assembled from scalar box integrals and tensor pentagon integrals. In the pure spinor representation
of \MafraGJA, this amounts to
\eqnn\UVC
 $$\eqalignno{
{\cal M}^{1-{\rm loop}}_4 \, \Big|_{\rm UV} &= {i  \over (4\pi)^{4} \epsilon} \big| \langle V_1 T_{2,3,4} \rangle \big|^2 = {i\over (4\pi)^{4} \epsilon} \big|s_{12} s_{23} A^{{\rm tree}}(1,2,3,4) \big|^2
 &\UVC\cr
{\cal M}^{1-{\rm loop}}_5 \, \Big|_{\rm UV} &= {i  \over (4\pi)^{4} \epsilon} \Big\{  \big| \langle V_1T^m_{2,3,4,5} \rangle  \big|^2+ \Big[ { \big| \langle V_{12} T_{3,4,5} \rangle \big|^2 \over s_{12} } + (2\leftrightarrow 3,4,5) \Big]  \cr
 & \ \ \ \ \ \ \  \ \ \ \ \ \ \  \ \ \ \ \ \ \ + \Big[ { \big| \langle V_{1} T_{23,4,5} \rangle \big|^2 \over s_{23} } + (2,3 |2,3,4,5) \Big]  \Big\}  \ .
 }$$
This is the low-energy limit of closed-string one-loop amplitudes, see \refs{\FiveSdual, \twoloopfive} for the
discussion of the five-point kinematic factor as well as \momker\ and \ininKLT\ for $S_0$ and $M_3$, respectively.
As discussed in section \UVRsym, the components with a $D$-dimensional dilaton and a graviton depend explicitly on the dimension $D$ via \dilatonD, and one gets
\eqnn\UVonel
$$\eqalignno{
&\big| \langle V_1T^m_{2,3,4,5} \rangle  \big|^2+ \Big[ { \big| \langle V_{12} T_{3,4,5} \rangle \big|^2 \over s_{12} } + (2\leftrightarrow 3,4,5) \Big] + \Big[ { \big| \langle V_{1} T_{23,4,5} \rangle \big|^2 \over s_{23} } + (2,3 |2,3,4,5) \Big]  \, \Big|^D_{{\rm IIB}}  \cr
&= \pmatrix{\tilde A^{\rm tree}(1,2,3,5,4)\cr
 \tilde A^{\rm tree}(1,3,2,5,4)}^{T} \cdot S_0 \cdot M_3 \cdot \pmatrix{ A^{\rm tree}(1,2,3,4,5)\cr A^{\rm tree}(1,3,2,4,5)}  
  \times 
  \cases{  \ \ \ \ \  1 \,  \ \ \ \ \ : \ \ h^5 \cr  \cr \displaystyle {(8-D)\over 6} \ : \ \phi h^4} 
  \ , &\UVonel
}$$
The factor $(8-D)/6$ in the $\phi h^4$ contribution implies that 
its five-point one-loop UV divergence in \UVC\ vanishes in the critical dimension 
and does not violate the R-symmetry of type IIB supergravity, see the corresponding 
discussion along with the two-loop UV divergence in section \UVRsym.

\listrefs

\bye